\documentclass[sigconf, nonacm]{acmart}

\usepackage{ifthen}

\newcommand{\version}{full} 

\usepackage{amsmath}
\usepackage{amsthm}
\usepackage{xcolor}
\usepackage{enumitem}
\setlist[itemize]{
    noitemsep,    
    topsep=0pt,   
    leftmargin=*, 
}
\setlist[enumerate]{
    noitemsep,    
    topsep=0pt,   
    leftmargin=*, 
}

\usepackage{caption}
\usepackage{adjustbox}
\usepackage{calc}
\usepackage{tikz}
\usetikzlibrary{decorations.pathreplacing,angles,quotes,positioning,shapes.misc,decorations.text}
\usepackage{subcaption}
\usepackage{xspace}
\usepackage{url}

\usepackage{wasysym}

\usepackage[normalem]{ulem}

\usepackage[ruled,vlined,noend,linesnumbered]{algorithm2e}

\usepackage{caption}

\usepackage{hyperref}

\usepackage{verbatimbox}

\usepackage{multirow}

\ifthenelse{\equal{\version}{conf}}{
  \newcommand{\inConfVersion}[1]{#1} 
  \newcommand{\inFullVersion}[1]{}   
}{
  \ifthenelse{\equal{\version}{full}}{
    \newcommand{\inConfVersion}[1]{} 
    \newcommand{\inFullVersion}[1]{#1} 
  }{
    \newcommand{\inConfVersion}[1]{#1} 
    \newcommand{\inFullVersion}[1]{#1} 
  }
}

\ifthenelse{\equal{\version}{conf}}{
  \usepackage[appendix=strip,bibliography=common]{apxproof}
}{
  \usepackage[appendix=append,bibliography=common]{apxproof}
}

\newtheoremrep{theorem}{Theorem}[section]
\newtheoremrep{proposition}[theorem]{Proposition}

\newcommand{\nameOfApproach}{{\sl RePMILA}\xspace}

\newcommand{\trewriting}{{promotion choices}\xspace}

\renewcommand{\epsilon}{\varepsilon}

\newcommand{\nop}[1]{}

\newcommand{\removespacetocaption}{\vspace{-0.3cm}}

\newcommand{\db}{\textbf{D}}
\newcommand{\type}[1]{\text{type}(#1)}
\newcommand{\schRel}{\textsf{Rels}}

\newcommand{\objectsRes}[1]{\objects_{#1}}

\newcommand{\Attr}[1]{\ensuremath{\text{Attr}(#1)}}
\newcommand{\myvar}[1]{\ensuremath{var}(#1)}

\newcommand{\eg}{\textit{e.g.}}

\newcommand{\trans}[1][i]{T_{#1}}

\newcommand{\transset}{{\mathcal{T}}}
\newcommand{\tfirst}{\textit{first}}
\newcommand{\objects}{\mathbf{Obj}}

\newcommand{\schedule}{s}
\newcommand{\schop}[1][\schedule]{O_{#1}}
\newcommand{\schord}[1][\schedule]{\leq_{#1}}
\newcommand{\schords}[1][\schedule]{<_{#1}}
\newcommand{\schvord}[1][\schedule]{\ll_{#1}}
\newcommand{\schvf}[1][\schedule]{v_{#1}}
\newcommand{\sstart}{\textit{op}_0}

\newcommand{\dependson}[3][\schedule]{#2 \rightarrow_{#1} #3}

\newcommand{\isolationlevel}{\text{$\mathcal{I}$}}
\newcommand{\mvrc}{\text{RC}\xspace}
\newcommand{\si}{\text{SI}\xspace}
\newcommand{\ssi}{\text{SSI}\xspace}

\newcommand{\rc}{\mvrc}
\newcommand{\alloc}{\mathcal{A}}
\newcommand{\allocb}{\mathcal{B}}
\newcommand{\allocrc}{\alloc_{\mvrc}}
\newcommand{\allocsi}{\alloc_{\si}}

\newcommand{\variables}{\mathbf{Var}}
\newcommand{\templ}[1][i]{\tau_{#1}}
\newcommand{\workload}{\mathcal{P}}
\newcommand{\templset}{\mathcal{P}}

\newcommand{\seg}[1]{SeG(#1)}
\newcommand{\cg}[1]{\seg{#1}}

\newcommand{\ptconflictgraph}{\text{pt-conflict-graph}}
\newcommand{\cyclesym}{\Gamma}

\newcommand{\prefix}[2]{{\normalfont\textsf{prefix}}_{#2}(#1)}

\newcommand{\postfix}[2]{{\normalfont\textsf{postfix}}_{#2}(#1)}

\newcommand{\isopostgres}{\ensuremath{\{\mvrc,\allowbreak\si,\allowbreak\ssi\}}\xspace}

\newcounter{conditioncounter}

\newcommand{\new}[1]{{#1}}

\newcommand{\myR}{\ensuremath{\mathtt{R}}}
\newcommand{\myW}{\ensuremath{\mathtt{W}}}
\newcommand{\myUP}{\ensuremath{\mathtt{U}}}
\newcommand{\R}[2][i]{\myR_{#1}\mathtt{[#2]}}
\newcommand{\W}[2][i]{\myW_{#1}\mathtt{[#2]}}
\newcommand{\UP}[2][i]{\myUP_{#1}\mathtt{[#2]}}
\newcommand{\CT}[1][i]{\mathtt{C}_{#1}}

\newcommand{\ReadSet}[1]{\ensuremath{\text{ReadSet}(#1)}}
\newcommand{\WriteSet}[1]{\ensuremath{\text{WriteSet}(#1)}}

\newcommand{\x}{\mathtt{t}}
\newcommand{\y}{\mathtt{v}}
\newcommand{\z}{\mathtt{q}}
\newcommand{\xx}{\mathtt{x}}
\newcommand{\yy}{\mathtt{y}}
\newcommand{\zz}{\mathtt{z}}
\newcommand{\X}{\mathtt{X}}
\newcommand{\Y}{\mathtt{Y}}
\newcommand{\Z}{\mathtt{Z}}

\newcommand{\vx}{\mathtt{X}}
\newcommand{\vy}{\mathtt{Y}}
\newcommand{\vz}{\mathtt{Z}}

\newcommand{\ListAttr}[1]{\ensuremath{\{\text{#1}\}}}
\newcommand{\Account}{\ensuremath{\text{Account}}\xspace}
\newcommand{\Savings}{\ensuremath{\text{Savings}}\xspace}
\newcommand{\Checking}{\ensuremath{\text{Checking}}\xspace}

\newcommand{\Balance}{\ensuremath{\text{Balance}}\xspace}
\newcommand{\DepositChecking}{\ensuremath{\text{DepositChecking}}\xspace}

\newcommand{\tmap}{\mu}

\newcommand{\seqpcq}{\mathcal{C}}

\newcommand{\ptrans}{transaction template}   
\newcommand{\ptranss}{transaction templates} 
\newcommand{\Ptranss}{Transaction templates} 
\newcommand{\PTranss}{Transaction Templates} 

\usetikzlibrary{tikzmark,scopes,decorations.pathreplacing}
\usetikzlibrary{arrows, arrows.meta}

\begin{document}

\title{Using Read Promotion and Mixed Isolation Levels for Performant Yet Serializable Execution of Transaction Programs}

\author{Brecht Vandevoort}
\affiliation{%
  \institution{UHasselt, Data Science Institute} 
}

\author{Alan Fekete}
\affiliation{%
  \institution{University of Sydney}
}

\author{Bas Ketsman}
\affiliation{%
\institution{Vrije Universiteit Brussel}
}

\author{Frank Neven}
\affiliation{%
  \institution{UHasselt, Data Science Institute} 
}

\author{Stijn Vansummeren}
\affiliation{%
  \institution{UHasselt, Data Science Institute} 
}

\begin{abstract}
    We propose a theory that can determine the lowest isolation level that can be allocated to each transaction program in an application in a mixed-isolation-level setting, to guarantee that all executions will be serializable and thus preserve all integrity constraints, even those that are not explicitly declared. This extends prior work applied to completely known transactions, to deal with the realistic situation where transactions are generated by running programs with parameters that are not known in advance. Using our theory, we propose an optimization method that allows for high throughput while ensuring that all executions are serializable. Our method is based on searching for application code modifications that are semantics-preserving while improving the isolation level allocation. We illustrate our approach to the SmallBank benchmark.
\end{abstract}



\maketitle

\section{Introduction}
\label{sec:introduction}

Transaction management is a core capability for database management systems. While research continues to find ways to improve performance, especially utilising novel hardware~\cite{DBLP:conf/sigmod/BernsteinDDP15,DBLP:conf/cidr/BernsteinRD11,DBLP:conf/sigmod/DiaconuFILMSVZ13,DBLP:conf/cloud/DingKDG15,DBLP:journals/pvldb/GuoCWQZ19,DBLP:journals/pvldb/HuangQKLS20,DBLP:conf/sigmod/JonesAM10,DBLP:conf/sigmod/KimWJP16,DBLP:journals/pvldb/LarsonBDFPZ11,DBLP:conf/sigmod/LimKA17,DBLP:conf/sigmod/0001MK15,DBLP:conf/sigmod/SharmaSD18}, the bulk of application software runs on popular platforms whose concurrency control mechanisms are decades old and are known to suffer from bottlenecks that make serializable transactions perform poorly under contention~\cite{vldbpaper,DBLP:journals/pvldb/YuBPDS14}. Under the narrative that many applications have domain-specific reasons why they do not need to be perfectly serializable, these platforms offer the application programmer a choice of isolation levels. As such, the programmer can select a weaker isolation level, such as the platform's default READ COMMITTED level~\cite{DBLP:journals/pvldb/BailisDFGHS13}, to improve performance when apt. However, there is not yet a well-grounded way for the programmer to decide when a decision to accept non-serializable isolation is justified.

\smallskip
\noindent
{\it Robustness of transactions to guarantee serializability.}
Recent theoretical studies have provided algorithms that can analyze the
collection of transactions occurring in an application, and determine which
isolation level to use for each of them in a mixed-isolation-level setting,
while still guaranteeing the robustness of the
application~\cite{DBLP:conf/pods/VandevoortKN23}. That is, every possible
execution of the application's transactions will in fact be serializable, even
though several of the transactions do not run with serializable isolation
level. This task, dubbed the \emph{allocation
  problem}~\cite{DBLP:conf/pods/Fekete05}, requires some characterisation of the
concurrency control mechanism used by the platform.  In particular, the
conclusions on a platform using traditional shared and exclusive locks operating
on single-version data will differ from the conclusions for multiversion systems
that allow reading versions that have been overwritten (and therefore do not
block reads)~\cite{DBLP:conf/pods/Fekete05,DBLP:journals/tods/KetsmanKNV22}. The
current theory for solving the allocation problem, however, assumes that all the
transactions are completely known at allocation time, including all the items
that will be read and written. This is not realistic: in practice applications
execute programs with parameters that are provided at run-time, after
allocation. For example, a student enrollment system will have a program that
enrolls a student to a particular course. The concrete student id and the course
code are only provided by the end-user when the program is executed.

\smallskip
\noindent
{\it Robustness of transaction templates.}
The key technical advance of this paper, is to provide an algorithm that can determine which isolation level can be allocated to each of a set of templates, where a template is an abstraction that aims to capture a transaction whose read and write set is determined based on some variables. The algorithm returns an allocation that is guaranteed to be robust on a platform such as PostgreSQL which offers multiversion concurrency control mechanisms. That is, every execution of any set of transactions that instantiate the templates with arbitrary values, will be serializable. 
%
\new{Our algorithm, therefore, supports any number of instantiations per template while maintaining polynomial time in the template size. Unlike earlier work~\cite{DBLP:conf/pods/VandevoortKN23}, which assumes a fixed set of concrete transactions and scales with their size, our execution time is independent of the number of concrete instantiations.}

The template abstraction we use also has some restrictions. It assumes a fixed set of read-only attributes that cannot be modified and are used to select tuples for updates. A common example of such attributes are primary keys. 
%
\new{This assumption prevents predicate reads that could cause non-serializable executions not covered by our analysis. While it excludes workloads like TPC-C which involve predicate reads, in many cases this is not a major limitation as 
the inability to update primary keys is not a significant limitation, because keys are typically assigned once and remain unchanged due to regulatory or data integrity requirements.}
Furthermore, as the template abstraction loses some of the constraints in the transaction code, our analysis is conservative but safe: we may miss a desirable allocation which is robust, but the allocation found by our algorithm does indeed ensure that all executions are serializable.
However, within the restrictions of the template abstraction, we can prove optimality of the allocation we identify, in the sense that we find the allocation that gives each template the lowest isolation level possible (i.e., prioritizing Read Committed over Snapshot Isolation, and Snapshot Isolation over the serializable level), such that any code that fits the template will be robust.
\new{While our template abstraction omits branches and loops for simplicity, we could handle them by treating each execution path as a separate template unfolding. Branches and bounded loops unfold easily, and for unbounded loops, it is sufficient to show that any non-robust behavior has a counterexample with loops unfolded only a fixed number of times (cf.~\cite{DBLP:conf/edbt/VandevoortK0N23}).}
\new{The current paper does not consider depedencies like foreign keys.
It is known that, in general, these dependencies can render the robustness property undecidable~\cite{DBLP:conf/icdt/VandevoortK0N22}.}

\smallskip
\noindent
{\it Optimization via read promotion and mixed-isolation level allocation.}
Our allocation algorithm can be used to find ways to deliver an application with the correctness guarantees of serializable execution (so all state invariants are preserved, including those which are not explicitly declared in the database), and yet better performance than when each transaction is executed at the serializable level. We propose to consider a space of different ways to modify the application code (while not changing its semantics), by ``promoting'' some read operations \cite{DBLP:journals/tods/FeketeLOOS05} so they are treated as identity updates, and thus set exclusive locks. For each of these different promotion choices, we use our allocation algorithm to determine the lowest isolation level. 
For each promotion choice and determined robust allocation we can then empirically measure the performance obtained; the promotion choice with best performance of its robust allocation, is how the application should be coded.
{We refer to this optimization approach as \emph{read promotion and mixed isolation level allocation (\nameOfApproach)}}.

We illustrate \nameOfApproach for the well-known benchmark SmallBank~\cite{Alomari:2008:CSP:1546682.1547288}, and explore the performance obtained on PostgreSQL under a range of workload parameters. We find that some promotion choices have a robust allocation whose throughput  is competitive with the throughput  of the unmodified application running  with all transactions using Read Committed. Unlike the latter, however, the robust allocation still guarantees serializable execution. Furthermore, the  throughput under robust allocation can be twice the throughput achieved by running all transactions under the platform's serializable isolation level.


\smallskip
\noindent
{\it Contributions.}
The contributions of this paper are varied, with both theory and empirical results, and we propose guidance for practitioners. As theory, we offer a proof technique that allows demonstrating robustness for an allocation of multi-version isolation levels for transaction templates, and a polynomial-time algorithm that generates a unique lowest robust allocation. We give an experimental demonstration for the SmallBank application mix, that some promotion choices allow performance of a robust allocation, close to that of the default non-robust allocation (and much better than the naive use of Serializable isolation for all transactions). We consider that in this context \nameOfApproach can be useful for practitioners as they seek performance while guaranteeing serializable execution.

\smallskip
\noindent
{\it Organization.}
The remainder of the paper is structured as follows. In Section~\ref{sec:SBexample} we illustrate \nameOfApproach applied to SmallBank, demonstrating how each program is abstracted as a template, how the allocation algorithm determines the allocation of isolation levels for the templates, how the various promotion choices are generated, and what allocation is generated for each promotion choice. In Section~\ref{sec:SBmeasured} we show the measured performance for the different promotion choices, and compare with baselines where all programs are run with Serializable isolation level, and also where all are run at default Read Committed isolation (and thus undeclared data invariants can be violated). Section~\ref{sec:theory} presents the theory and includes the details of the allocation algorithm. We discuss related work in Section~\ref{sec:relwork}. Finally, Section~\ref{sec:discussion} looks at implications and limitations of this work, and identifies some further research directions.

\inConfVersion{A full version of this paper is available as \cite{fullversion_repmila}, containing all proofs, \new{additional examples, further intuition regarding formalization}, as well as the SQL code for the SmallBank benchmark.}


\section{Read promotion and mixed isolation level allocation (\nameOfApproach)}
\label{sec:SBexample}



To explain our approach, we work through the way it is applied to the well-known SmallBank benchmark application~\cite{Alomari:2008:CSP:1546682.1547288}. While it is not a real-world example, it has enough features to illustrate how we can identify a high-performing, robust allocation, and yet it is simple enough to fit in a conference paper.

\smallskip
\noindent
{\it SmallBank benchmark.}
The SmallBank~\cite{Alomari:2008:CSP:1546682.1547288} schema consists of the tables
Account(\underline{Name}, CustomerID), Savings(\underline{CustomerID}, Balance), and
Checking(\uline{Cus\-tomerID}, Balance) (key attributes are underlined).
The \Account table associates customer names with IDs. The other tables contain the balance (numeric value) of the savings and checking accounts of customers identified by their ID. 
The application code interacts with the database via 
transaction programs:
\begin{itemize}
\item
Balance($N$) returns the total balance (savings and checking) for a customer with name $N$.

\item
DepositChecking($N$,$V$) makes a deposit of amount $V$ 
in the checking account of the customer with name $N$ (see Figure~\ref{fig:intro:depositchecking}).

\item
TransactSavings($N$,$V$) makes a deposit or withdrawal $V$ on the savings account of the customer with name $N$.

\item
Amalgamate($N_1$,$N_2$) transfers all the funds from customer $N_1$ to customer $N_2$.

\item
Finally, WriteCheck($N$,$V$) writes a check $V$ against the account of the customer with name $N$, penalizing if overdrawing.
\end{itemize}

\smallskip
\noindent
{\it Transaction templates.} We abstract transaction programs via transactions templates as illustrated in 
Figure~\ref{fig:smallbank-abstract-syntax}.
\inFullVersion{The corresponding SQL code is provided in 
Section~\ref{sec:app:smallbank} in the appendix.}
A \ptrans{} consists of a sequence of read ($\texttt{R}$), write ($\texttt{W})$, and update ($\texttt{U}$) operations. Each operation accesses exactly one tuple.
 For instance, $\R[]{\vx :\Account\{N,C\}\}}$ indicates that a read operation is performed to a tuple $\vx$ in relation $\Account$ on the attributes Name and CustomerID. We abbreviate the names of attributes by their first letter to save space. The set $\{N,C\}$ is the read set of the read operation.
Similarly, $\myW$ and $\myUP$ refer to write and update operations to tuples of a specific relation. Write operations have an associated write set while update operations contain a read set followed by a write set: e.g.,
$\UP[]{\vz :\DepositChecking\{C,B\}\{B\}\}}$ first reads the CustomerID and Balance of tuple $\vz$ and then writes to the attribute Balance.
A $\myUP$-operation is an atomic update that first reads the tuple and then writes to it. Templates serve as abstractions of transaction programs and represent an infinite number of possible workloads. 
For instance, let  $\xx$, $\yy$, $\zz$ (and their primed versions) 
be concrete database objects serving as interpretations of the variables
$\X$, $\Y$, and $\Z$. Then, 
disregarding attribute sets, $\{ R[\xx]R[\yy]R[\zz]\allowbreak U[\zz], R[\xx']R[\yy']R[\zz']U[\zz']$,$R[\xx]\allowbreak U[\zz]\}$ is a workload consistent with the SmallBank templates as it contains two instantiations of Write\-Check and one instantiation of DepositChecking. We remark that $\{R[\xx]R[\yy]\allowbreak R[\zz]\allowbreak U[\zz']\}$
with $\zz\ne\zz'$ is not a valid workload as the two final operations in WriteCheck should be on the same object as required by the formalization.
Typed variables effectively enforce domain constraints as we assume that variables that range over tuples of different relations can never be instantiated by the same value. For instance, in the \ptrans{} for \DepositChecking in Figure~\ref{fig:smallbank-abstract-syntax}, $\vx$ and $\vz$ can not be interpreted to be the same object.

\begin{figure}[t]
    \begin{minipage}[c]{0.99\columnwidth}
    %
    \begin{verbbox}[\footnotesize]
    DepositChecking(N,V):
      SELECT CustomerId INTO :X FROM Account WHERE Name=:N;
      UPDATE Checking SET Balance = Balance+:V
        WHERE CustomerId=:X;
      COMMIT;
    \end{verbbox}
    
    \begin{center}
    {
        \theverbbox
    }
    \end{center}
    
    \removespacetocaption
    
        \caption{SQL code for \DepositChecking.}
        \label{fig:intro:depositchecking}
    
    
    
    \end{minipage}%
    
    \vspace{.5em}
    \begin{minipage}[c]{0.99\columnwidth}
        \centering\footnotesize
    
    \begin{minipage}[t]{0.5\textwidth-2ex}
    \Balance: 
    \[
    \begin{array}{l}
    \R[]{\vx: \Account\ListAttr{N, C}}\\
    \R[]{\vy: \Savings\ListAttr{C, B}}\\
    \R[]{\vz: \Checking\ListAttr{C, B}}\\
    \end{array}
    \]
    DepositChecking: 
    \[
    \begin{array}{l}
    \R[]{\vx: \Account\ListAttr{N, C}}\\
    \UP[]{\vz: \Checking\ListAttr{C, B}\ListAttr{B}}\\
    \end{array}
    \]
    TransactSavings: 
    \[
    \begin{array}{l}
    \R[]{\vx: \Account\ListAttr{N, C}}\\
    \UP[]{\vy: \Savings\ListAttr{C, B}\ListAttr{B}}\\
    \end{array}
    \]
    \end{minipage}%
    \hfill%
    \begin{minipage}[t]{0.50\textwidth-2ex}
    Amalgamate: 
    \[
    \begin{array}{l}
    \R[]{\vx_1: \Account\ListAttr{N, C}}\\
    \R[]{\vx_2: \Account\ListAttr{N, C}}\\
    \UP[]{\vy_1: \Savings\ListAttr{C, B}\ListAttr{B}}\\
    \UP[]{\vz_1: \Checking\ListAttr{C, B}\ListAttr{B}}\\
    \UP[]{\vz_2: \Checking\ListAttr{C, B}\ListAttr{B}}\\
    \end{array}
    \]
    WriteCheck: 
    \[
    \begin{array}{l}
    \R[]{\vx: \Account\ListAttr{N, C}}\\
    \R[]{\vy: \Savings\ListAttr{C, B}}\\
    \R[]{\vz: \Checking\ListAttr{C, B}}\\
    \UP[]{\vz: \Checking\ListAttr{C, B}\ListAttr{B}}\\
    \end{array}
    \]
    \end{minipage}
    
    
        \caption{Transaction templates for SmallBank. 
        }
        \label{fig:smallbank-abstract-syntax}
    
    \end{minipage}
\end{figure}
Templates do not capture all constraints in the original programs, and may therefore overapproximate the transactions that can occur when the actual programs are executed. For instance, the workload $\{ R[\x]U[\z], R[\x]U[\z']\}$ is consistent with the SmallBank templates (two instantiations of DepositChecking), but cannot occur in practice under the assumption that a customer can only have one checking account.

\enlargethispage{+4mm}

\vspace{0.1cm}
\noindent
{\it Lowest robust allocation.} We are interested in determining the lowest isolation level for each separate template such that every execution that arises under the assigned levels will in fact be serializable. We  refer to such as an allocation as \emph{robust}. We consider the isolation levels of PostgreSQL: Read Committed (RC), Snapshot Isolation (SI), and Serializable Snapshot Isolation (SSI) where we rank them from \emph{lower} to \emph{higher} as RC $<$ SI $<$ SSI,  under the assumption that throughput increases when isolation levels are lowered. The allocation algorithm that we describe in Section~\ref{sec:theory} finds that the allocation that maps DepositChecking to RC and all other templates to SSI is robust, and is in fact optimal in the sense that no isolation level can be lowered without breaking robustness.

\vspace{0.1cm}
\noindent
{\it Read promotion choices.}
None of the considered isolation levels allow dirty writes. This  forces a transaction 
that wants to overwrite a change made by an earlier transaction, to wait until the earlier transaction either commits or aborts. Therefore, if we promote a read operation to an update (that is, a read operation that writes back the observed value), the semantics of the transaction remains unaffected but the lowest robust allocation might differ.   
Ignoring the read operations over the read-only Account table,
the SmallBank benchmark contains 4 read operations over the Savings and Checking relations that are candidates for promotion, resulting in 16 possible promotion choices.
For each promotion choice, we run our algorithm to detect the lowest robust allocation. The resulting allocations are summarized in Table~\ref{tab:smallbank:detected}.
We denote each promotion choice by the read operations that are promoted. For example, \textit{`Bal: S, WC: C'} promotes the read operation over the Savings relation in the Balance program, and the read operation over the Checking relation in the WriteCheck program. For convenience, the promotion choices are grouped by their lowest robust allocation.
Notice that without promotion, as mentioned previously, only one out of the five programs (nl., DepositChecking) can be executed under an isolation level lower that SSI without giving up serializability. Furthermore, introducing a few promoted reads quickly leads to robust allocations where almost all programs are being executed under RC. 
However, the best promotion choice is not necessarily the one that allows the most programs to run under RC, for the simple reason that the newly introduced writes could have a negative impact on the overall performance. Throughput experiments are therefore needed to determine the best promotion choices, as we discuss in the next section.

\begin{table}
    \begin{center}\small
    \begin{tabular}{rc|cccccl}
         & \textbf{Promotion} & \multicolumn{5}{c}{\textbf{Lowest robust allocation}} & \\
        & \textbf{choices}& \textbf{Bal} & \textbf{DC} & \textbf{TS} & \textbf{Am} & \textbf{WC} & \\
        \hline
        (1) & {no promotion} & \multirow{2}{*}{SSI} & \multirow{2}{*}{RC} & \multirow{2}{*}{SSI} & \multirow{2}{*}{SSI} & \multirow{2}{*}{SSI} & \multirow{2}{*}{(A)}\\
        (2) & WC: C &  &  &  &  &  \\
        \hline
        (3) & Bal: S & \multirow{2}{*}{SSI} & \multirow{2}{*}{SSI} & \multirow{2}{*}{SSI} & \multirow{2}{*}{SSI} & \multirow{2}{*}{SSI} & \multirow{2}{*}{(B)} \\
        (4) & Bal: S, WC: C &  &  &  &  &  \\
        \hline
        (5) & {Bal: C} & \multirow{4}{*}{SI} & \multirow{4}{*}{RC} & \multirow{4}{*}{RC} & \multirow{4}{*}{RC} & \multirow{4}{*}{SI} & \multirow{4}{*}{(C)}\\
        (6) & {WC: S} &  &  &  &  &  \\
        (7) & Bal: C, WC: S &  &  &  &  &  \\
        (8) & Bal: C, WC: C &  &  &  &  &  \\
        \hline
        (9) & {WC: S,C} & \multirow{2}{*}{SI} & \multirow{2}{*}{RC} & \multirow{2}{*}{RC} & \multirow{2}{*}{RC} & \multirow{2}{*}{RC} & \multirow{2}{*}{(D)} \\
        (10) & Bal: C, WC: S,C &  &  &  &  &  \\
        \hline
        (11) & {Bal: S,C} & \multirow{4}{*}{RC} & \multirow{4}{*}{RC} & \multirow{4}{*}{RC} & \multirow{4}{*}{RC} & \multirow{4}{*}{SI} & \multirow{4}{*}{(E)}\\
        (12) & {Bal: S, WC: S} &  &  &  &  &  \\
        (13) & Bal: S,C, WC: S &  &  &  &  &  \\
        (14) & Bal: S,C, WC: C &  &  &  &  &  \\
        \hline
        (15) & {Bal: S, WC: S,C} & \multirow{2}{*}{RC} & \multirow{2}{*}{RC} & \multirow{2}{*}{RC} & \multirow{2}{*}{RC} & \multirow{2}{*}{RC} & \multirow{2}{*}{(F)} \\
        (16) & Bal: S,C, WC: S,C &  &  &  &  & 
    \end{tabular}
    \end{center}

    \caption{
    Lowest robust allocations for each \trewriting over the SmallBank benchmark, grouped by allocation. \new{Promotion choices and allocations are labeled for easy reference.}}
    \label{tab:smallbank:detected}
\end{table}

\section{Evaluating \nameOfApproach over SmallBank}
\label{sec:SBmeasured}


Here we show experimentally the performance achieved by our approach when  applied to the SmallBank benchmark. 

\smallskip
\noindent
{\it Experimental setup.}
All experiments use PostgreSQL 16.2 as the database engine, running on a single machine with two 18-core Xeon Gold 6240 CPUs (2.6 GHz), 192 GB RAM and 200GB local SSD storage. A separate machine is used to query the database, with 100 concurrent clients executing randomly sampled programs from the SmallBank benchmark. If the database aborts a transaction, the client immediately retries the same program with the same parameters until it successfully commits. Each experiment runs for 60 seconds and is repeated 5 times to measure the average throughput.
The database is populated with 18000 accounts. A small subset of 20 accounts act as a hotspot that will be accessed more frequently. The level of contention is varied by changing the probability of sampling a hotspot account during execution. Within the hotspot, uniform sampling is used to select an account. Unless otherwise specified, each of the five SmallBank programs has an equal probability of being sampled by the clients. Throughput is indicated in number of transactions per second. 
We implemented the allocation algorithm described in Section~\ref{sec:theory} in Python. Verifying whether an allocation is robust takes only a few seconds, whereas computing the lowest allocation for a specific promotion choice requires {less than a minute}. This runtime is acceptable since the computation is performed only once and can be executed offline.

\begin{figure}
    \centering
    \includegraphics[width=\linewidth]{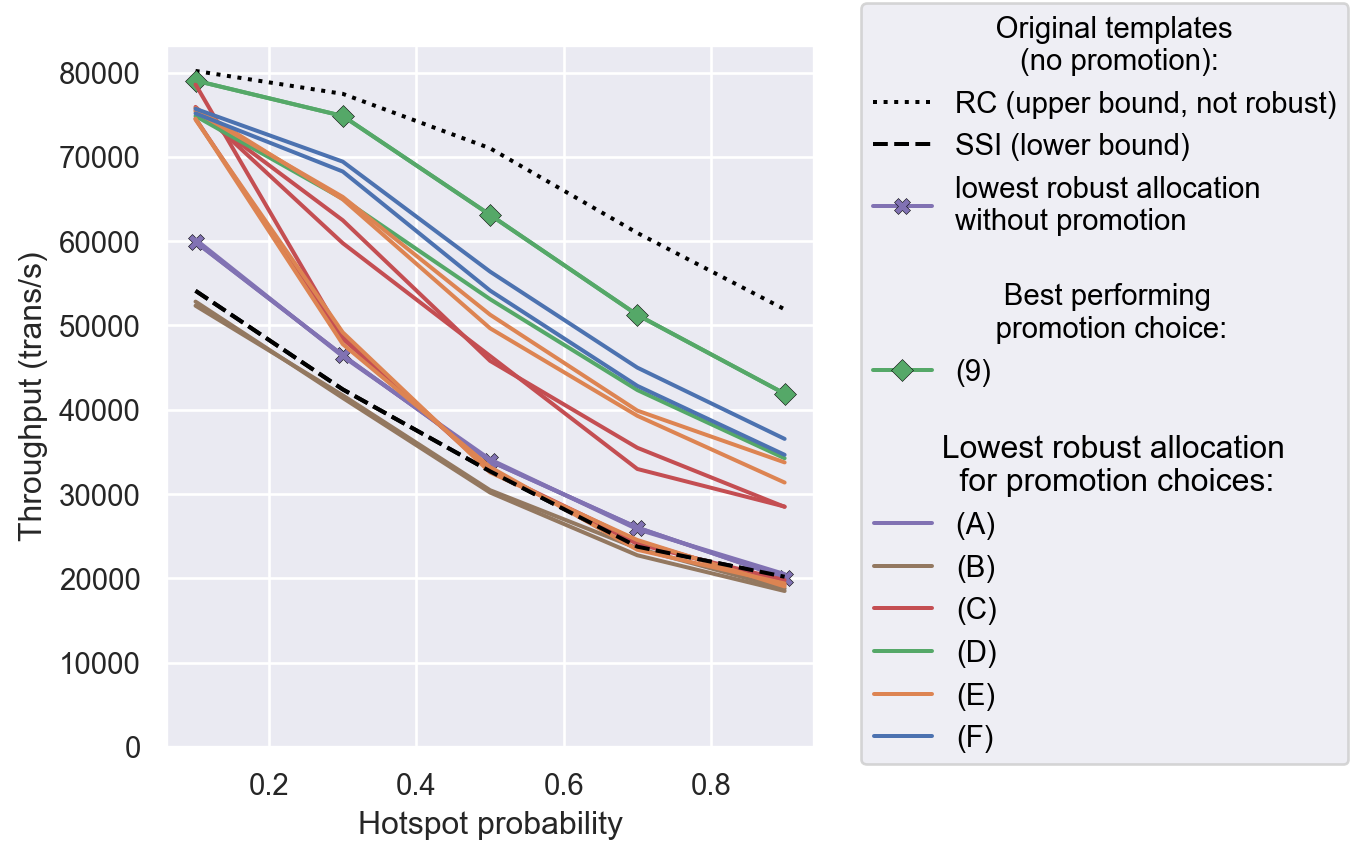}
    \caption{Throughput for the promotion choices mentioned in Table~\ref{tab:smallbank:detected} for various hotspot probabilities.} 
       
    \label{fig:sb:prob}
\end{figure}

\inFullVersion{
\begin{figure}
    \centering
    \includegraphics[width=\linewidth]{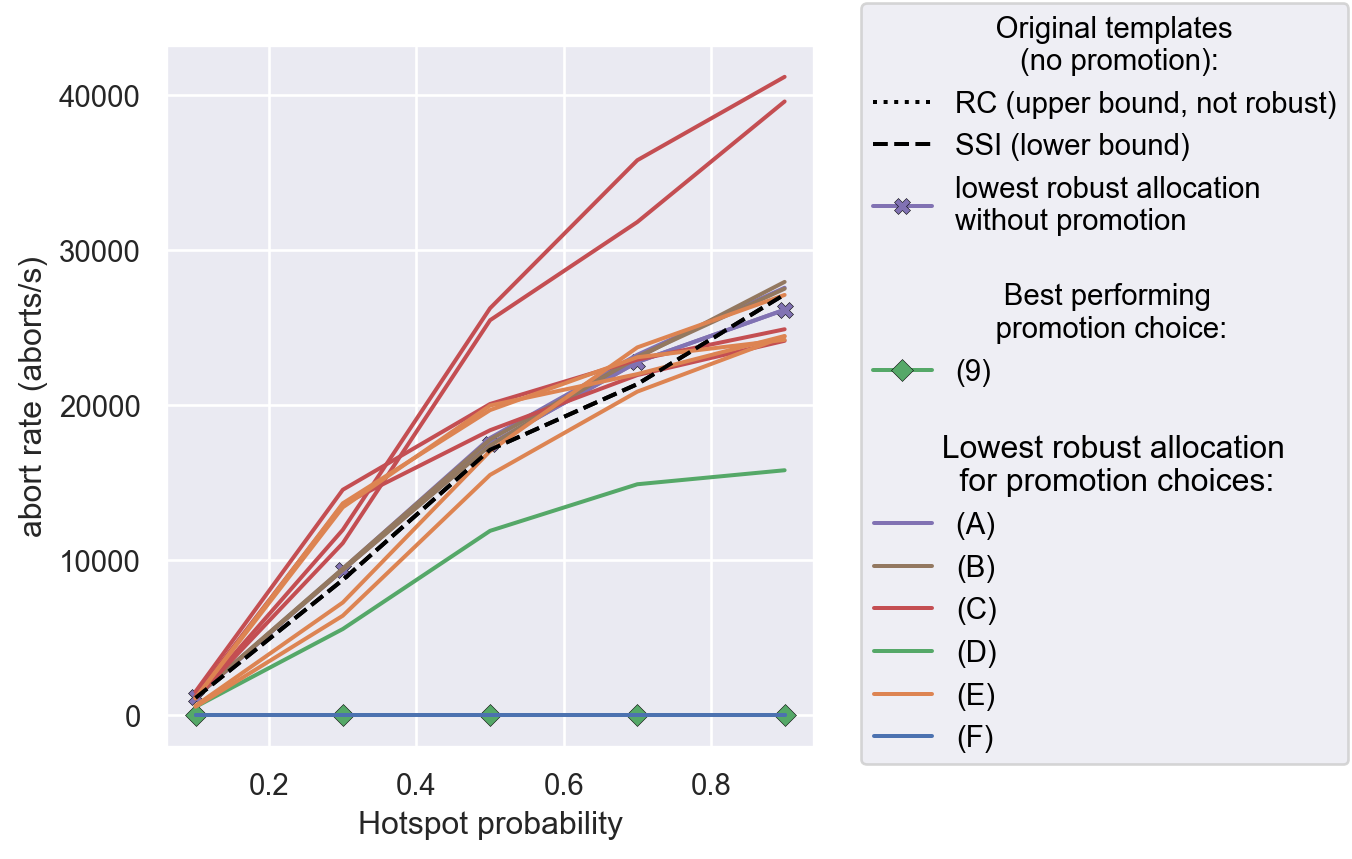}
    \caption{\new{Abort rate for the promotion choices mentioned in Table~\ref{tab:smallbank:detected} for various hotspot probabilities.}} 
       
    \label{fig:sb:prob:abort}
\end{figure}
}


\subsection{\nameOfApproach improves performance} 
 Figure~\ref{fig:sb:prob} displays the throughput for the different promotion choices and associated lowest robust allocations mentioned in Table~\ref{tab:smallbank:detected} for increasing levels of contention. Promotion choices that result in the same lowest robust allocation are depicted in the same color and are not individually labeled to avoid adding complexity to the figure.
 While not all lines are easily discernible, we do see that the throughput for almost all promotion choices is higher than executing the unmodified templates under SSI.  The exception is the bottom line indicating that these choices perform worse than executing the original unmodified programs under SSI. 
 This is due to the introduction of additional writes, which do not provide allocation benefits as all templates still require the use of SSI.
 
 None of the promotions can match the throughput levels that can be reached by the nonrobust allocation that executes all unmodified templates at RC. Nevertheless, the most performant promotion choice \textit{`WC: S,C'} is a near match. Here, no reads are promoted in Balance and the lowest robust allocation assigns SI to Balance and RC to the others. In our experiments we observe that in this case there are no aborts due to concurrent writes. 

We point out that the lowest robust allocation for the unmodified templates allows little to no improvement compared to running everything under SSI, especially when contention increases. We thus conclude that read promotions can increase throughput
and that considering various promotion choices is helpful. In general, read promotions that allow to allocate RC tend to outperform those requiring SI or SSI. However, there are some notable exceptions: for example, \textit{`WC: S,C'} still requires SI for Balance, but it outperforms all promotion choices that only require RC. Similarly, there are promotion choices (which are not discernible in the figure as individual lines are not labeled) where the lowest allocation assigns both Balance and WriteCheck to SI, yet these choices still outperform specific promotion choices that allocate only WriteCheck to SI, and RC to all others.

\inFullVersion{
\new{
    Figure~\ref{fig:sb:prob:abort} shows the abort rates corresponding to each experiment in Figure~\ref{fig:sb:prob}.
    As expected, we notice an increased number of aborts whenever the hotspot probability increases. We observe that the three best performing choices in Figure~\ref{fig:sb:prob} lead to settings without aborts. It is furthermore interesting to note that although these three choices have no aborts, there is still a notable difference in throughput, with the choice with fewer promoted reads (i.e., promotion choice (9)) outperforming the others.
}
}

\subsection{Lowest allocation outperforms higher ones}
\begin{figure}
    \centering
    \includegraphics[width=\linewidth]{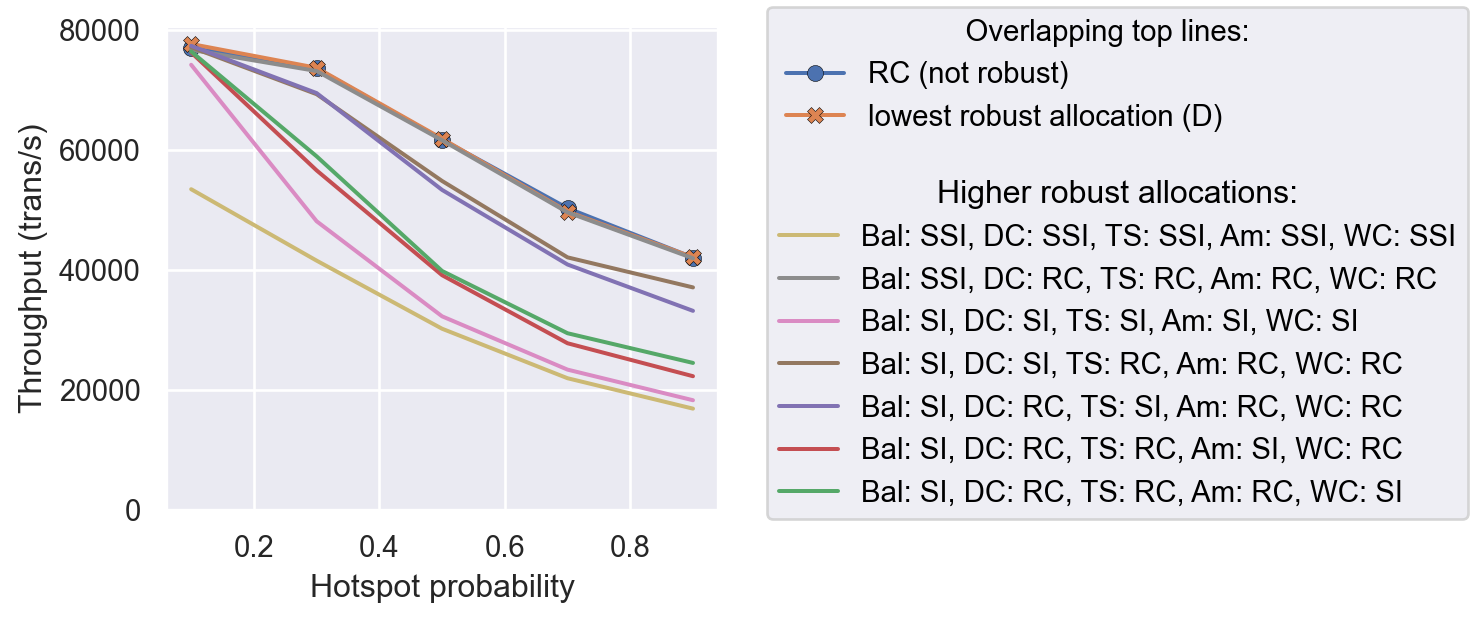}
    \caption{Throughput for promotion choice \new{(9)} `WC: S,C' under its 
    lowest 
    and higher allocations. 
    }
    \label{fig:sb:optimalalloc} 
\end{figure}

We defined a robust allocation as \emph{lowest} when no isolation level can be reduced without compromising robustness, following the ordering RC $<$ SI $<$ SSI. As shown in Section~\ref{sec:theory}, there always exists a \emph{unique} lowest allocation with respect to this order. We verify that these unique lowest allocation consistently outperforms higher allocations, allowing us to focus solely on them.  
To this end, we examine the most performant promotion choice, \textit{`WC: S,C'}, and compare its lowest robust allocation—where Balance is assigned SI and all other templates receive RC—against all alternative allocations that raise the isolation level for at least one of the five programs. Figure~\ref{fig:sb:optimalalloc} demonstrates that the lowest robust allocation indeed outperforms all higher allocations. Additionally, we observe that its throughput matches that of the nonrobust allocation for \textit{`WC: S,C'}, where all templates run under RC.  

\inFullVersion{
    \new{Figure~\ref{fig:sb:optimalalloc:abort} visualizes the corresponding abort rates. The unique lowest allocation results in no aborts, whereas the abort rate for the higher allocations quickly increases for higher hotspot probabilities.}
}

\inFullVersion{
\begin{figure}
    \centering
    \includegraphics[width=\linewidth]{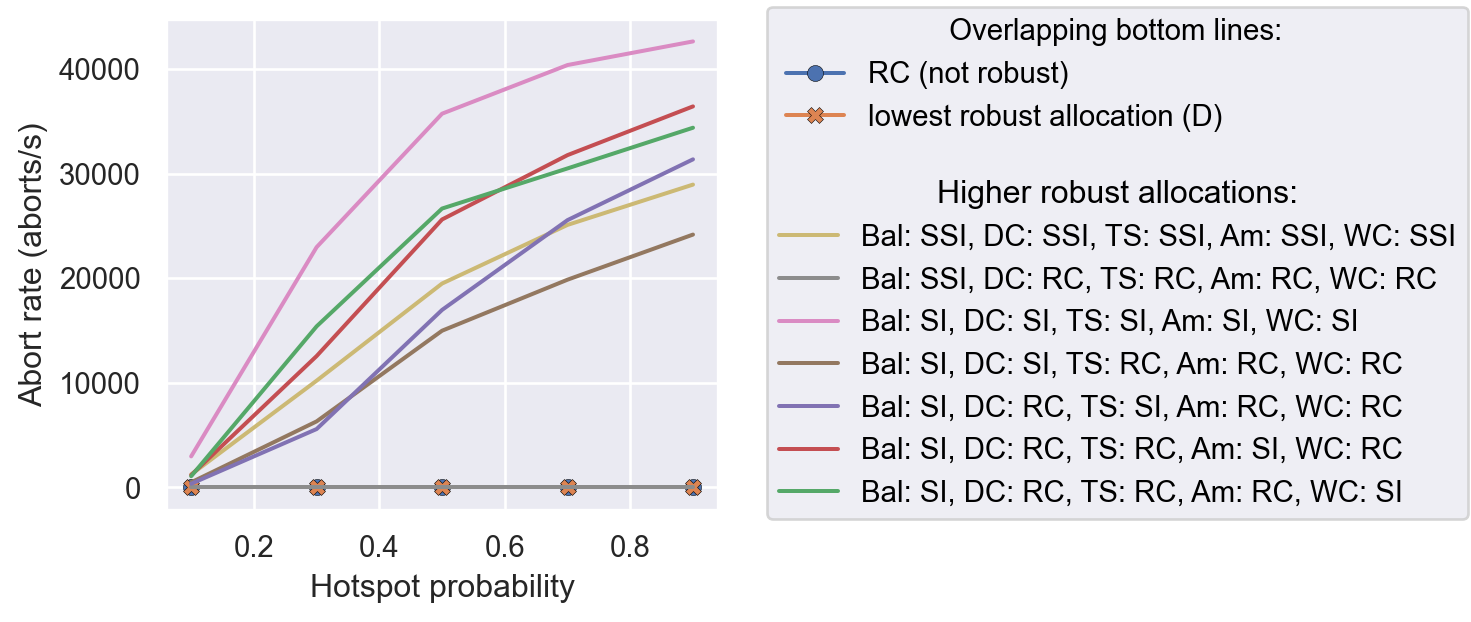}
    \caption{\new{Abort rate for promotion choice (9) `WC: S,C' under its 
    lowest 
    and higher allocations. 
    }}
    \label{fig:sb:optimalalloc:abort} 
\end{figure}
}

\begin{figure}
    \centering
    \includegraphics[width=\linewidth]{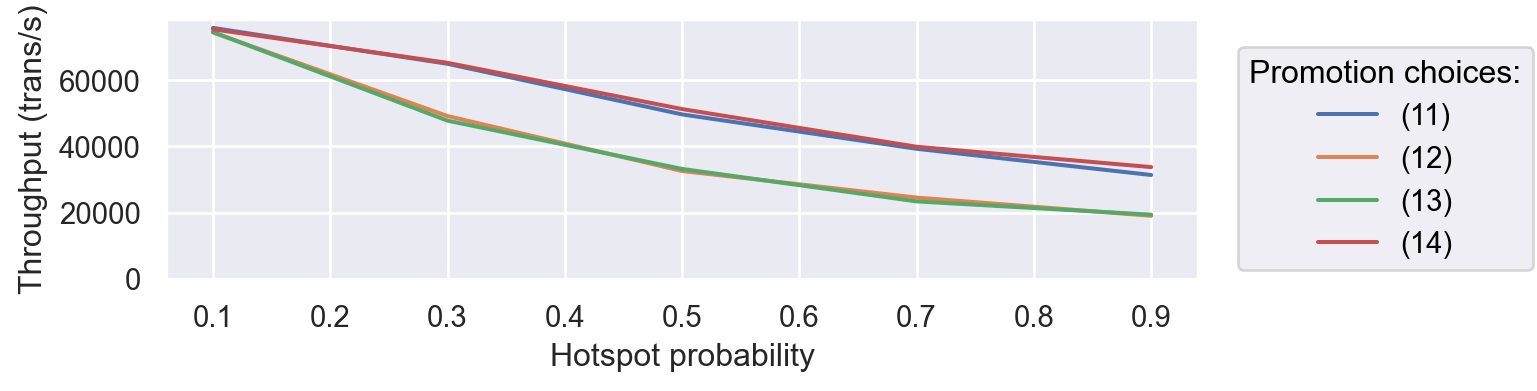}
    \caption{Throughput for different promotion choices sharing the same lowest robust allocation \new{(E)}, i.e., the one that maps WriteCheck to SI and all the others to RC.} 
    \label{fig:sb:optimalstrategy}
\end{figure}

\inFullVersion{
\begin{figure}
    \centering
    \includegraphics[width=\linewidth]{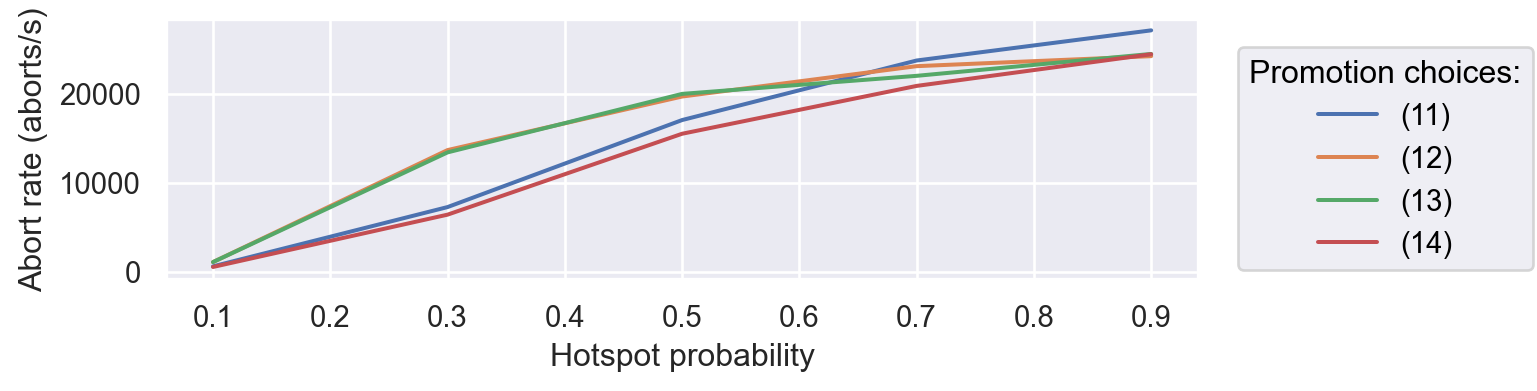}
    \caption{\new{Abort rate for different promotion choices sharing the same lowest robust allocation {(E)}, i.e., the one that maps WriteCheck to SI and all the others to RC.}} 
    \label{fig:sb:optimalstrategy:abort}
\end{figure}
}

\begin{figure}[h]
    \centering
    \includegraphics[width=\linewidth]{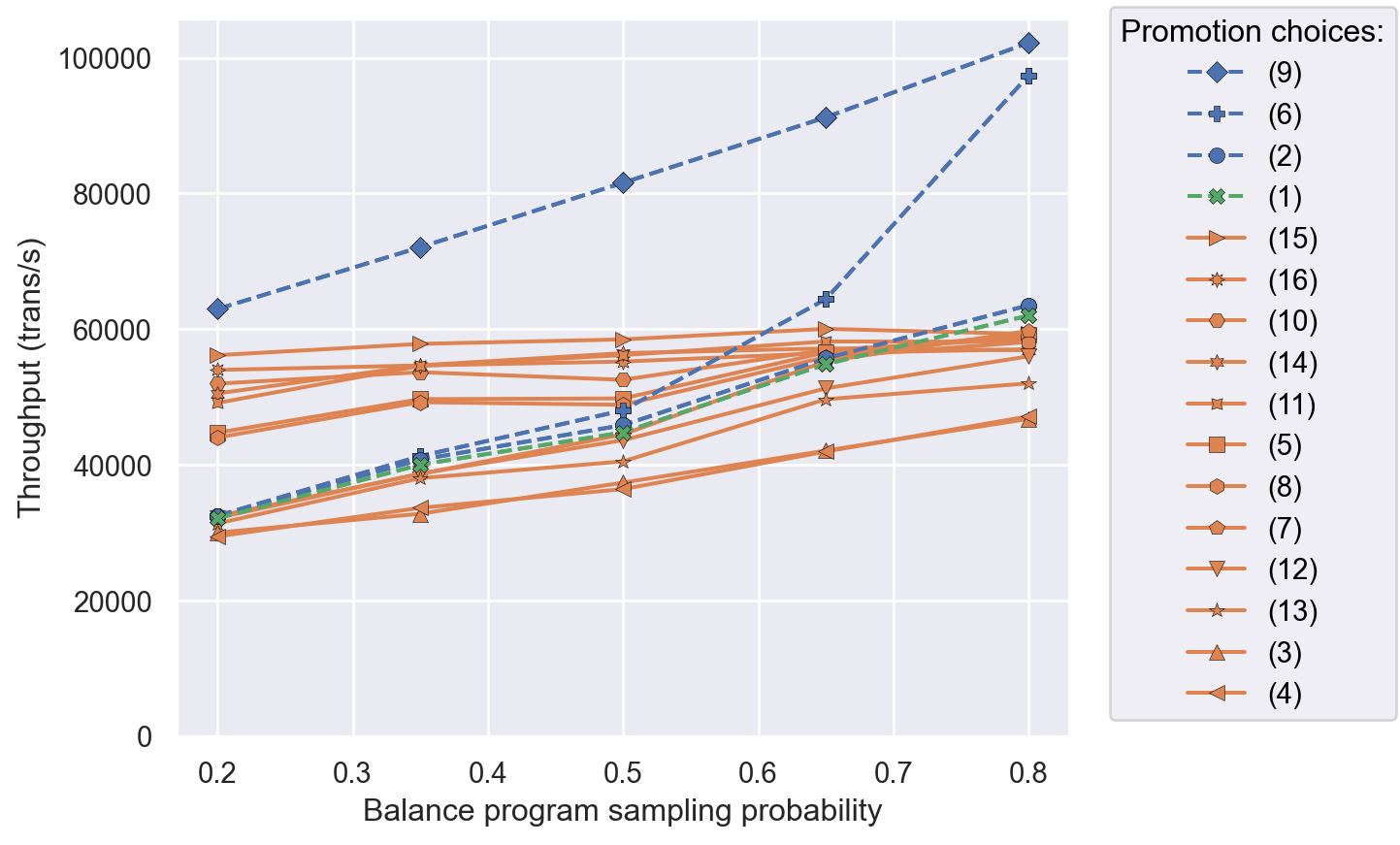}
    \caption{Throughput for different promotion choices varying the probability of executing Balance for a fixed hotspot probability of 0.5. Promotion choices that do not promote a read in Balance are in {\color{blue}blue}. Those that do are in {\color{orange}orange}.} 
    \label{fig:sb:readonly}
\end{figure}

\inFullVersion{
\begin{figure}
    \centering
    \includegraphics[width=\linewidth]{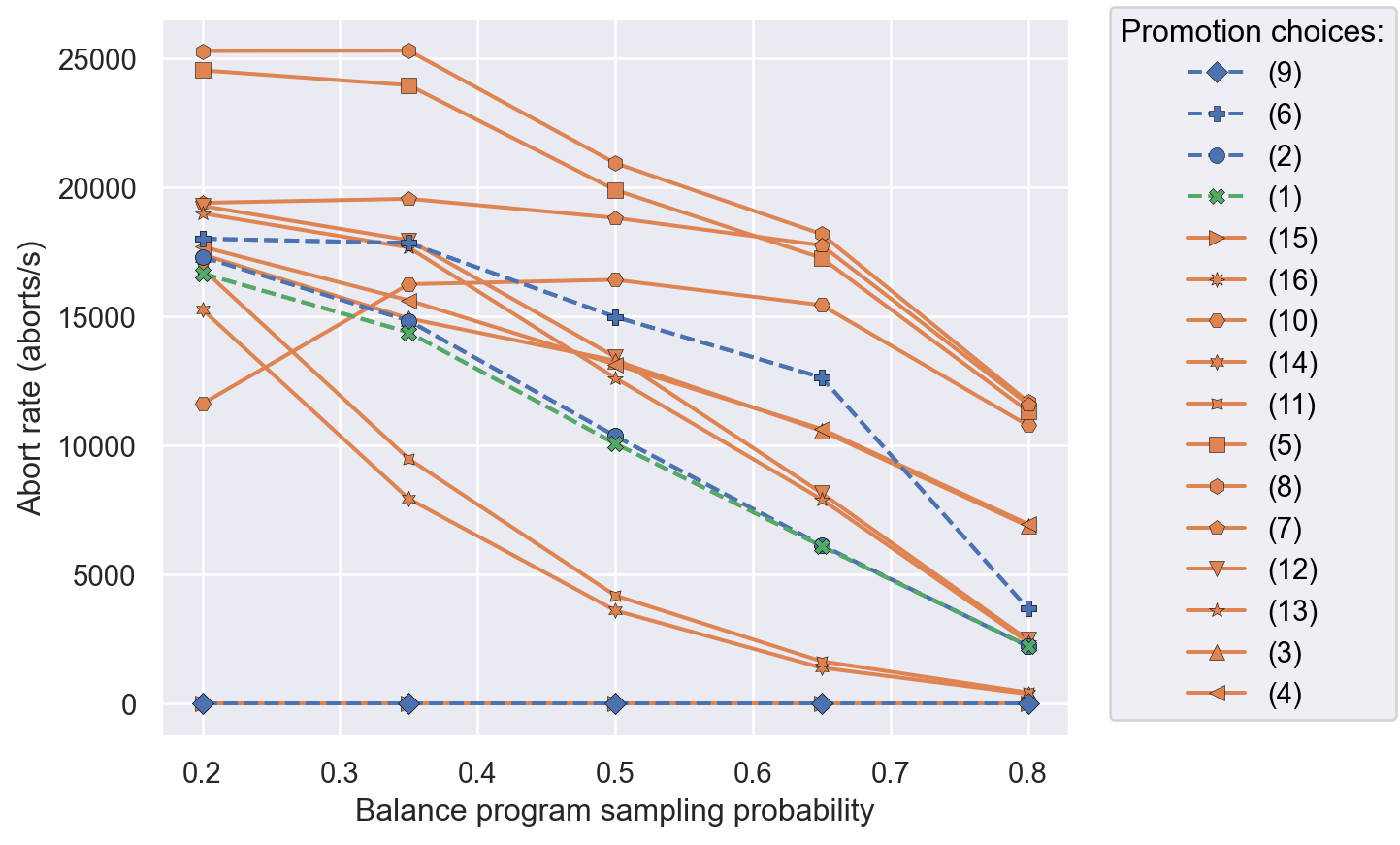}
    \caption{\new{Abort rate for different promotion choices varying the probability of executing Balance for a fixed hotspot probability of 0.5. Promotion choices that do not promote a read in Balance are in {\color{blue}blue}. Those that do are in {\color{orange}orange}.}} 
    \label{fig:sb:readonly:abort}
\end{figure}
}


\subsection{Promotion choices within the same lowest robust allocation impact performance}

As illustrated by Table~\ref{tab:smallbank:detected}, distinct promotion choices may lead to the same lowest robust allocation. For example, the four promotion choices \textit{`Bal: S,C'}, \textit{`Bal: S, WC: S'}, \textit{`Bal: S,C, WC: S'} and \textit{`Bal: S,C, WC: C'} all result in the same lowest robust allocation that maps WriteCheck to SI and all the others to RC.
Since these promotion choices promote different reads, an identical lowest allocation does not necessarily lead to identical performance even under lower levels of contention, as is shown in Figure~\ref{fig:sb:optimalstrategy}. There, we see that \textit{`Bal: S,C'} and \textit{`Bal: S,C, WC: C'} greatly outperform the others. 
This implies, in particular, that an experimental exploration of promotion choices cannot be limited to just a single promotion choice per unique robust allocation. A closer inspection reveals that the read promotions of the top performer, \textit{`Bal: S,C, WC: C'}, form a strict \emph{superset} of those of the next-best contender, \textit{`Bal: S,C'}. 
In this case, the slight performance gain is obtained by taking an earlier lock due to the newly promoted read. Indeed, the promoted read in WriteCheck is over a tuple written to by a later write in the same transaction. Since the lock is taken earlier and since WriteCheck uses SI, potentially concurrent writes are detected earlier, thereby avoiding the amount of work wasted before an abort.
\inFullVersion{
    \new{
        The corresponding abort rates are shown in Figure~\ref{fig:sb:optimalstrategy:abort}. From this Figure, we can see that there is only a small difference in abort rates when the hotspot probability is high, even though the throughput difference is still significant.
    }
}



\subsection{Impact of template frequency}

We next illustrate that finding the best performing promotion choices is influenced as well by the template frequency. The previous experiments assumed uniform sampling over the five possible SmallBank programs. Since only Balance is read-only (assuming no promoted reads), this corresponds to a workload where only 20\% of the transactions is read-only. To further explore the impact of promoting reads in read-only transactions, Figure~\ref{fig:sb:readonly} shows the throughput for different promotion choices when the probability of executing Balance is varied between 0.2 and 0.8, assuming a fixed hotspot probability of 0.5. The remaining four templates are sampled with equal probability. Overall, we see that promotion choices that do not promote a read in Balance start to outperform those that do when the probability of executing Balance increases. This observation is most pronounced for the promotion choice \textit{`WC: S'}, which is among the worst performers when the probability of executing Balance is low, but becomes the second best performer when the probability is high, vastly outperforming all other promotion choices. Closer inspection reveals that in the latter case even the original programs (i.e., `no promotion') outperform all promotion choices that promote a read in Balance.
\inFullVersion{
    \new{
        The corresponding abort rates are shown in Figure~\ref{fig:sb:readonly:abort}. For high probabilities of executing Balance, we see that the abort rates for promotion choices that do not promote a read in Balance drop significantly. Although some other promotion choices also have low abort rates in this case, they do not perform as well in terms of throughput. This is most likely because the reduced concurrency due to additional write locks results in a low throughput, not outweighing reduced number of aborts.
    }
}



\subsection{Discussion}
\label{sec:eval:discussion}

The SmallBank exploration in this section validates \nameOfApproach\ as an effective optimization method, demonstrating that combining read promotion with the lowest robust allocation can double throughput compared to executing all transactions under SSI. Furthermore, it achieves throughput comparable to the unsafe yet default RC isolation level used by some platforms, all while preserving safety.

\new{Since our performance gains stem from a tradeoff between reduced concurrency and fewer aborts inherent to the studied isolation levels, we expect our results to generalize to other systems supporting these levels. While the analysis in Table~\ref{tab:smallbank:detected} should hold, the optimal promotion choice may vary (cf.\ Alomari et al.\cite{Alomari:2008:CSP:1546682.1547288} for cross-system comparisons).}

\new{In theory, the number of promotion choices grows exponentially with the total number of read operations in templates. However, the number of required throughput experiments can be significantly reduced by the following guidelines: \emph{(i)} ignoring reads from read-only tables (e.g., Accounts in SmallBank); \emph{(ii)} avoiding promotion in read-only templates, particularly when they frequently appear in workloads (e.g., Balance); and, \emph{(iii)} when multiple promotion choices yield the same lowest allocation, prioritizing those that promote reads occurring earlier in a template.}


\section{Allocation algorithm}
\label{sec:theory}

We start by introducing all necessary terminology and borrow notation from~\cite{DBLP:conf/pods/VandevoortKN23,vldbpaper} along with some examples that we adapt and modify to suit our context. In particular, we discuss transactions, schedules, conflict-serializability, and isolation levels in Section~
\ref{sec:trans-schedules}--\ref{sec:isolevels}. In Section~\ref{sec:templates}--\ref{sec:def-robustness} we introduce templates and their robustness. Finally, we 
present an algorithm for template robustness in Section~\ref{sec:robustness} and an algorithm for finding the lowest robust allocation in Section~\ref{sec:allocation}.

\subsection{Transactions and Schedules}
\label{sec:trans-schedules}

\newcommand{\oldexample}[1]{{\color{gray}#1}}

A \emph{relational schema} is a set $\schRel$ of relation names. For
each $R\in\schRel$, $\Attr{R}$ is the finite set of associated attribute names and 
we fix an infinite set $\objectsRes{R}$ of abstract objects called tuples.
We assume that $\objectsRes{R} \cap \objectsRes{S} = \emptyset$ for all  $R,S\in\schRel$ with $R\neq S$. We denote by $\objects$ the set $\bigcup_{R\in\schRel} \objectsRes{R}$ of all possible tuples. We require that for every $\x\in\objects$ there is a unique relation $R \in \schRel$ such that $\x\in\objectsRes{R}$. 
We then say that $\x$ is of \emph{type} $R$ and denote the latter by $\type{\x}=R$. 
A \emph{database} $\db$ over schema $\schRel$ assigns to every relation name $R\in \schRel$ a finite set $R^{\db} \subset\objectsRes{R}$.

For a tuple $\x \in \objects$, we distinguish three operations $\R[]{\x}$, $\W[]{\x}$, and $\UP[]{\x}$ on $\x$, denoting that $\x$ is read, written, or updated, respectively. {We say that the operation is \emph{on} the tuple $\x$.}
Here, $\UP[]{\x}$ is an atomic update and should be viewed as an atomic sequence of a read of $\x$ followed by a write to $\x$. 
To differentiate between the cases where we want to refer to an actual operation (\myR, \myW, or \myUP) or to operations with a specific property (read or write), 
we employ the following terminology. A \emph{read operation} is an $\R[]{\x}$ or a $\UP[]{\x}$, and a \emph{write operation} is a $\W[]{\x}$ or a $\UP[]{\x}$. Furthermore, an \myR-operation is an $\R[]{\x}$, a \myW-operation is a $\W[]{\x}$, and a \myUP-operation is a $\UP[]{\x}$.
We also assume a special \emph{commit} operation denoted $\CT[]$.
{To every operation $o$ on a tuple of type $R$, we associate the set of attributes
$\ReadSet{o}\subseteq\Attr{R}$ and $\WriteSet{o}\subseteq\Attr{R}$ containing, respectively, the set of attributes that $o$ reads from and writes to. When $o$ is an $\myR$-operation then $\WriteSet{o}=\emptyset$. Similarly, when $o$ is a $\myW$-operation then $\ReadSet{o}=\emptyset$. 
} 

A \emph{transaction} $\trans[]$ is a sequence of read and write operations followed by a commit. 
Formally, we model a transaction as a linear order $(\trans[],\leq_{\trans[]})$, where $\trans[]$ is the set of (read, write and commit) operations occurring in the transaction and $\leq_{\trans[]}$ encodes the ordering of the operations. As usual, we use $<_{\trans[]}$ to denote the strict ordering. We denote the first operation in $T$ by $\tfirst(T)$.

When considering a set $\transset$ of transactions, we assume that every transaction in the set has a unique id $i$ and write $\trans$ to make this id explicit. Similarly, to distinguish the operations of different transactions, we add this id as a subscript to the operation. {That is, we write $\W{\x}$, $\R{\x}$, and $\UP{\x}$ to denote a $\W[]{\x}$, $\R[]{\x}$, and $\UP[]{\x}$ occurring in transaction $\trans$; similarly $\CT[i]$ denotes the commit operation in transaction $T_i$. }
This convention is consistent with the literature (see, \eg\
\cite{DBLP:conf/sigmod/BerensonBGMOO95,DBLP:conf/pods/Fekete05}). 
To avoid ambiguity of notation, we assume that a transaction performs at most one write, one read, and one update per tuple.
The latter is a common assumption (see, \eg~\cite{DBLP:conf/pods/Fekete05}). All our results carry over to the more general setting in which multiple writes and reads per tuple are allowed.

A \emph{(multiversion) schedule} $\schedule$ over a set $\transset$ of transactions is a tuple $(\schop, \schord, \schvord, \schvf)$ where
\begin{itemize}
    \item $\schop$ is the set containing all operations of transactions in $\transset$ as well as a special operation $\sstart$ conceptually writing the initial versions of all existing tuples,
    \item $\schord$ encodes the ordering of these operations,
    \item $\schvord$ is a \emph{version order} providing for each tuple $\x$ a total order over all write operations on $\x$ occurring in $\schedule$, and,
    \item $\schvf$ is a \emph{version function} mapping each read operation $a$ in $\schedule$ to either $\sstart$ or to a write operation different from $a$ in $\schedule$ (recall that a write operation is either a $\W[]{x}$ or a $\UP[]{x}$).
\end{itemize}
We require that $\sstart \leq_\schedule a$ for every operation $a \in {O_\schedule}$, {$\sstart \ll_\schedule a$ for every write operation $a \in {O_\schedule}$}, and that $a <_{\trans[]} b$ implies $a <_\schedule b$ for every $\trans[] \in \transset$ and every $a,b \in \trans[]$. 
We furthermore require that for every read operation $a$, $v_\schedule(a) <_\schedule a$ and, if $v_\schedule(a) \neq \sstart$, then the operation $v_\schedule(a)$ is on the same tuple as $a$.
Intuitively, $\sstart$ indicates the start of the schedule, the order of operations in $s$ is consistent with the order of operations in every transaction $\trans[]\in\transset$, and the version function maps each read operation $a$ to the operation that wrote the version observed by $a$.
If $v_\schedule(a)$ is $\sstart$, then $a$ observes the initial version of this tuple.
The version order $\ll_\schedule$ represents the order in which different versions of a tuple are installed in the database. For a pair of write operations on the same tuple, this version order does not necessarily coincide with $\leq_\schedule$.
E.g., under \mvrc and \si the version order is based on the commit order instead.

Figure~\ref{fig:ex:schedule} depicts an example of a schedule. There, the read operations on $\x$ in $\trans[1]$ and $\trans[4]$ both read the initial version of $\x$ instead of the version written but not yet committed by $\trans[2]$. Furthermore, the read operation $\R[2]{\y}$ in $\trans[2]$ reads the initial version of $\y$ instead of the version written by $\trans[3]$, even though $\trans[3]$ commits before $\R[2]{\y}$.

We say that a schedule $\schedule$ is a \emph{single version schedule} if {$\schvord$ is compatible with $\schord$ and} every read operation always reads the last written version of the tuple. Formally, {for each pair of write operations $a$ and $b$ on the same tuple, $a \schvord b$ iff $a \schords b$, and} for every read operation $a$ there is no write operation $c$ on the same tuple as $a$ with $v_\schedule(a) \schords c  \schords a$.
A single version schedule over a set of transactions $\transset$ is \emph{single version serial} if its transactions are not interleaved with operations from other transactions. That is, for every $a,b,c \in {O_\schedule}$ with $a <_{\schedule}
b<_{\schedule} c$ and $a,c \in \trans[]$ implies $b \in \trans[]$ for every
$\trans[] \in \transset$.

\inFullVersion{The absence of aborts in our definition of schedule is consistent with the common assumption~\cite{DBLP:conf/pods/Fekete05,DBLP:conf/concur/0002G16} that an underlying recovery mechanism will rollback aborted transactions. We consider isolation levels that only read committed versions. Therefore there will never be cascading aborts.}

\subsection{Conflict-Serializability}
\label{sec:ser}

Let $a_j$ and $b_i$ be two operations on the same tuple from different transactions $\trans[j]$ and $\trans[i]$ in a {set of transactions $\transset$}. We then say that {$a_j$ is \emph{conflicting} with $b_i$} if:
\begin{itemize}
    \item \emph{(ww-conflict)} $\WriteSet{a_j} \cap \WriteSet{b_i} \neq \emptyset$; or,
    \item \emph{(wr-conflict)} $\WriteSet{a_j} \cap \ReadSet{b_i} \neq \emptyset$; or, 
    \item \emph{(rw-conflict)} $\ReadSet{a_j} \cap \WriteSet{b_i} \neq \emptyset$.
\end{itemize}
{We also say that $a_j$ and $b_i$ are conflicting operations.}
Commit operations and the special operation $\sstart$ never conflict with any other operation.
When $a_j$ and $b_i$ are conflicting operations in $\transset$, we say that $a_j$ \emph{depends on} $b_i$ in a schedule $\schedule$ over $\transset$, denoted $b_i \rightarrow_\schedule a_j$ if:\footnote{Throughout the paper, we adopt the following convention:  a $b$ operation can be understood as a `before' while an $a$ can be interpreted as an `after'.}
\begin{itemize}
    \item \emph{(ww-dependency)} {{$b_i$ is ww-conflicting with $a_j$} and $b_i \ll_{\schedule} a_j$}; or, 
    \item \emph{(wr-dependency)} {{$b_i$ is wr-conflicting with $a_j$} and $b_i = v_\schedule(a_j)$ or $b_i \ll_{\schedule} v_\schedule(a_j)$}; or, 
    \item \emph{(rw-antidependency)} {{$b_i$ is rw-conflicting with $a_j$} and $v_\schedule(b_i) \ll_{\schedule} a_j$.}
\end{itemize}

Intuitively, a ww-dependency from $b_i$ to $a_j$ implies that $a_j$ writes a version of a tuple {that is installed} after the version written by $b_i$.
A wr-dependency from $b_i$ to $a_j$ implies that $b_i$ either writes the version observed by $a_j$, or it writes a version that is {installed} before the version observed by $a_j$.
A rw-antidependency from $b_i$ to $a_j$ implies that $b_i$ observes a version {installed} before the version written by $a_j$.

For example, the dependencies $\dependson[]{{\UP[2]{\x}}}{\W[4]{\x}}$, $\dependson[]{{\UP[3]{\y}}}{\R[4]{\y}}$ and $\dependson[]{\R[4]{\x}}{{\UP[2]{\x}}}$ are respectively a ww-dependency, a wr-dependency and a rw-antidependency in schedule $\schedule$ presented in Figure~\ref{fig:ex:schedule}.

Two schedules $\schedule$ and $\schedule'$ are \emph{conflict-equivalent} if they are over the same set $\transset$ of transactions and for every pair of conflicting operations $a_j$ and $b_i$, $\dependson[\schedule]{b_i}{a_j}$ iff $\dependson[\schedule']{b_i}{a_j}$.

\begin{definition}
    A schedule $\schedule$ is \emph{conflict-serializable} if it is conflict-equivalent to a single version serial schedule.
\end{definition}

A {\emph{serialization graph}} $\cg{\schedule}$ for schedule $\schedule$ over a set of transactions $\transset$ is the graph whose nodes are the transactions in $\transset$ and where there is an edge from $T_i$ to $T_j$ if {$T_j$ has an operation $a_j$ that depends on an operation $b_i$ in $T_i$, thus with $\dependson{b_i}{a_j}$.}
Since we are usually not only interested in the existence of
    {dependencies between operations}, but also in the operations themselves, we assume the
existence of a labeling function $\lambda$ mapping each edge to a set of pairs
of operations. Formally, $(b_i,a_j)\in \lambda(T_i,T_j)$ iff there is an
operation $a_j\in T_j$ that depends on an operation $b_i\in T_i$.
For ease of notation, we choose to
represent $\cg{\schedule}$ as a set of quadruples $(T_i,b_i,a_j,T_j)$ denoting
all possible pairs of these transactions $T_i$ and $T_j$ with all possible
choices of operations {with $\dependson{b_i}{a_j}$}. Henceforth, we refer to these quadruples simply as edges. Notice that edges cannot contain commit operations.

A \emph{cycle} $\cyclesym$ in $\cg{\schedule}$ is a non-empty sequence of edges $$(T_1,b_1,a_2,T_2),(T_2,b_2,a_3,T_3),\ldots, (T_n,b_n,a_1,T_1)$$ in $\cg{\schedule}$, in which every transaction is mentioned exactly twice. Note that cycles are by definition simple. Here, transaction $T_1$ starts and concludes the cycle. For a transaction $T_i$ in $\cyclesym$, we denote by $\cyclesym[T_i]$ the cycle obtained from $\cyclesym$ by letting $T_i$ start and conclude the cycle while otherwise respecting the order of transactions in $\cyclesym$. That is, $\cyclesym[T_i]$ is the sequence
$$
    (T_i,b_i,a_{i+1},T_{i+1})\cdots (T_n,b_n,a_1,T_1)(T_1,b_1,a_2,T_2)
    \cdots  (T_{i-1},b_{i-1},a_i,T_i).
$$

\begin{theorem}[implied by \cite{DBLP:conf/icde/AdyaLO00}]\label{theo:not-conflict-serializable}
    A schedule $\schedule$ is conflict-serializable iff $\cg{\schedule}$ is acyclic.
\end{theorem}


\inFullVersion{Figure~\ref{fig:ex:sergraph} visualizes the serialization graph $\cg{\schedule}$ for the schedule $\schedule$ in Figure~\ref{fig:ex:schedule}. Since $\cg{\schedule}$ is not acyclic, $\schedule$ is not conflict-serializable.}


Our formalisation of transactions and conflict serializability is based on \cite{DBLP:conf/pods/Fekete05}, generalized to operations over attributes of tuples and extended with $\myUP$-operations that combine $\myR$- and $\myW$-operations into one atomic operation. These definitions are closely related to the formalization presented by Adya et al.~\cite{DBLP:conf/icde/AdyaLO00}, but we assume a total rather than a partial order over the operations in a schedule. 


\subsection{Isolation Levels}
\label{sec:isolevels}

Let $\isolationlevel$ be a class of isolation levels.
An \emph{$\isolationlevel$-allocation $\alloc$} for a set of transactions $\transset$ is a function mapping each transaction $\trans[] \in \transset$ onto an isolation level $\alloc(\trans[]) \in \isolationlevel$. When $\isolationlevel$ is not important or clear from the context, we sometimes also say allocation rather than $\isolationlevel$-allocation. In this paper, we consider the following isolation levels: read committed (\mvrc), snapshot isolation (\si), and serializable snapshot isolation (\ssi). So, in general, $\isolationlevel=\isopostgres$.
Before we define what it means for a schedule to consist of transactions adhering to different isolation levels, we introduce some necessary terminology.
Some of these notions are illustrated in Example~\ref{ex:schedule} below.

\begin{figure}

    \begin{center}

        \begin{tikzpicture}[remember picture]
            \node[anchor=west](node0) at (0,0.5) {$\sstart$};
            \node[anchor=west] at (0,0)
            {$\phantom{\sstart\,\W[2]{\x}\,\R[4]{y}\,
                        \W[3]{\x}\,\CT[3]}\, \subnode{node11}{\R[1]{\x}}\, \CT[1]\, \phantom{\R[2]{\y}\, \W[2]{\z}\, \CT[2]\, \W[3]{\z}\, \CT[3]}$};
            \node[anchor=west] at (0,-0.5) {$\phantom{\sstart}\,
                    \subnode{node21}{{\UP[2]{\x}}}\,\phantom{\R[4]{\y}\, \W[3]{\y}\, \CT[3]\, \R[1]{\x}\, \CT[1]}\,
                    \subnode{node22}{\R[2]{\y}}\, \CT[2]$};
            \node[anchor=west] at (0,-1)
            {$\phantom{\sstart\,\W[2]{\x}\,\R[4]{\y}}\,
                    \subnode{node31}{{\UP[3]{\y}}}\,
                    \CT[3]$};
            \node[anchor=west] at (0,-1.5)
            {$\phantom{\sstart\,\W[2]{\x}}\,\subnode{node41}{\R[4]{\x}}\,
                    \phantom{\W[3]{\y}\,
                        \CT[3]\, \R[1]{\x}\, \CT[1]\, \R[2]{\y}\, \CT[2]}\,
                    \subnode{node42}{\W[4]{\x}}\,\subnode{node43}{\R[4]{\y}}\,
                    \CT[4]$};
            \draw[->,solid,out=180,in=12] (node11) to (node0);
            \draw[->,solid,in=-15,out=180] (node22) to (node0);
            \draw[->,-{Triangle[open,length=1.5mm,width=1.5mm]},solid,double,in=150,out=-100] (node0) to (node21);
            \draw[->,solid,in=-135,out=180] (node41) to (node0);
            \draw[->,-{Triangle[open,length=1.5mm,width=1.5mm]},solid,double,out=-45,in=135] (node0) to  (node31);
            \draw[->,-{Triangle[open,length=1.5mm,width=1.5mm]},solid,double,in=180,out=-35] (node21) to (node42);
            {\draw[->,solid,in=-70,out=120] (node21) to (node0);}
            {\draw[->,solid,out=110,in=-25] (node31) to (node0);}
            \coordinate[above left=15pt of node43] (dx);
            \coordinate[below=4pt of node22] (dy);
            \draw[->,solid] (node43) to[out=140,in=0] (dy) to[out=180,in=40] (node31);
            \node at (-0.5,0) {$\trans[1]:$};
            \node at (-0.5,-0.5) {$\trans[2]:$};
            \node at (-0.5,-1) {$\trans[3]:$};
            \node at (-0.5,-1.5) {$\trans[4]:$};
        \end{tikzpicture}

    \end{center}

    \caption{\label{fig:ex:schedule}
        A schedule $\schedule$ with $\schvf$ (single lines) and $\schvord$ (double lines) represented through
        arrows.
    }

\end{figure}

\inFullVersion{
\begin{figure}
    \begin{center}
        \begin{tikzpicture}
            \node[draw,circle] (T1) at (-3,-0.5) {$\trans[1]$};
            \node[draw,circle] (T2) at (2.5,0.5) {$\trans[2]$};
            \node[draw,circle] (T3) at (4.5,0.5) {$\trans[3]$};
            \node[draw,circle] (T4) at (-1,-0.5) {$\trans[4]$};
            \draw[->,bend left] (T1) edge[""] node [anchor=east,xshift=-15pt,yshift=-10pt,fill=white,inner sep=1pt]{\scalebox{0.7}{\ensuremath{\{(\R[1]{\x},\UP[2]{\x})\}}}} (T2);
            \draw[->,bend right] (T1) edge[""] node [anchor=north,yshift=-1pt,inner sep=1pt] {\scalebox{0.7}{\ensuremath{\{(\R[1]{\x},\W[4]{\x})\}}}} (T4);
            \draw[->,bend left] (T2) edge[""] node[anchor=center,fill=white,inner sep=1pt]{\scalebox{0.7}{\ensuremath{\{(\UP[2]{\x},\W[4]{\x})\}}}} (T4);
            \draw[->,bend left] (T4) edge[""] node[inner sep=1pt,anchor=center,fill=white]{\scalebox{0.7}{\ensuremath{\{(\R[4]{\x},\UP[2]{\x})\}}}} (T2);
            \draw[->,bend left] (T2) edge[""] node[inner sep=1pt,anchor=south,yshift=1pt,fill=white]{\scalebox{0.7}{\ensuremath{\{(\R[2]{\y},\UP[3]{\y})\}}}} (T3);
            \draw[->,bend left] (T3) edge[""] node [inner sep=1pt,anchor=west,xshift=15pt,yshift=10pt,fill=white]{\scalebox{0.7}{\ensuremath{\{(\UP[3]{\y},\R[4]{\y})\}}}} (T4);
        \end{tikzpicture}
    \end{center}
    \caption{\label{fig:ex:sergraph}
        {Serialization graph $\cg{\schedule}$ for the schedule $\schedule$ presented in Figure~\ref{fig:ex:schedule}.}
    }
\end{figure}
}

Let $\schedule$ be a schedule for a set $\transset$ of transactions.
Two transactions $\trans[i], \trans[j] \in \transset$ are said to be \emph{concurrent} in $\schedule$ when their execution overlaps. That is,
if $\tfirst(\trans[i]) \schords \CT[j]$ and $\tfirst(\trans[j]) \schords \CT[i]$.
We say that a write operation {$o_j$ on $\x$} in a transaction $\trans[j] \in \transset$ \emph{respects the commit order of $\schedule$} if the version of $\x$ written by $\trans[j]$ is installed after all versions of $\x$ installed by transactions committing before $\trans[j]$ commits, but before all versions of $\x$ installed by transactions committing after $\trans[j]$ commits. More formally, if for every write operation {$o_i$ on $\x$} in a transaction $\trans[i] \in \transset$ different from $\trans[j]$ we have {$o_j \schvord o_i$} iff $\CT[j] \schords \CT[i]$.
We next define when a read operation $a\in \trans[]$ reads the last committed version relative to a specific operation. For \rc this operation is $a$ itself while for \si this operation is $\tfirst(\trans[])$.
    {Intuitively, these definitions enforce that read operations in transactions allowed under \rc act as if they observe a snapshot taken right before the read operation itself, while under \si they observe a snapshot taken right before the first operation of the transaction.}
A read operation {$o_j$ on $\x$} in a transaction $\trans[j] \in \transset$ is \emph{read-last-committed in $\schedule$ relative to an operation $a_j \in \trans[j]$} (not necessarily different from {$o_j$}) if the following holds:
\begin{itemize}
    \item {$\schvf(o_j) = \sstart$} or $\CT[i] \schords a_j$ with $\schvf(o_j)\in \trans[i]$; and
    \item there is no write operation {$o_k$ on $\x$ in $\trans[k]$} with $\CT[k] \schords a_j$ and {$\schvf(o_j) \schvord o_k$}.
\end{itemize}
The first condition says that {$o_j$} either reads the initial version or a committed version, while the second condition states that
{$o_j$} observes the most recently committed version of $\x$ (according to $\schvord$). 
A transaction $\trans[j] \in \transset$ \emph{exhibits a concurrent write in $\schedule$} if there {is another transaction $\trans[i]\in\transset$ and there} are two write operations $b_i$ and $a_j$ in $\schedule$ on the same object with $b_i \in \trans[i]$, $a_j \in \trans[j]$ and $\trans[i] \neq \trans[j]$ such that $b_i \schords a_j$ and $\tfirst(\trans[j]) \schords \CT[i]$. That is, transaction $\trans[j]$ writes to an object that has been modified earlier by a concurrent transaction $\trans[i]$.

A transaction $\trans[j] \in \transset$ \emph{exhibits a dirty write in $\schedule$} if there are two write operations $b_i$ and $a_j$ in $\schedule$ on the same object with $b_i \in \trans[i]$, $a_j \in \trans[j]$ and $\trans[i] \neq \trans[j]$ such that $b_i \schords a_j \schords \CT[i]$.
That is, transaction $\trans[j]$ writes to an object that has been modified earlier by $\trans[i]$, but $\trans[i]$ has not yet issued a commit.
Notice that by definition a transaction exhibiting a dirty write always exhibits a concurrent write.
Transaction $\trans[4]$ in Figure~\ref{fig:ex:schedule} exhibits a concurrent write, since it writes to $\x$, which has been modified earlier by a concurrent transaction $\trans[2]$. However, $\trans[4]$ does not exhibit a dirty write, since $\trans[2]$ has already committed before $\trans[4]$ writes to $\x$.

\begin{definition}
    Let $\schedule$ be a schedule over a set of transactions $\transset$.
    A transaction $\trans[i] \in \transset$ is \emph{allowed under isolation level read committed (\mvrc) in $\schedule$} if:
    \begin{itemize}
        \item each write operation in $\trans[i]$ respects the commit order of $\schedule$;
        \item each read operation $b_i \in \trans[i]$ is read-last-committed in $\schedule$ relative to $b_i$; and
        \item $\trans[i]$ does not exhibit dirty writes in $\schedule$.
    \end{itemize}
    A transaction $\trans[i] \in \transset$ is \emph{allowed under isolation level snapshot isolation (\si) in $\schedule$} if:
    \begin{itemize}
        \item each write operation in $\trans[i]$ respects the commit order of $\schedule$;
        \item each read operation in $\trans[i]$ is read-last-committed in $\schedule$ relative to $\tfirst(\trans[i])$; and
        \item $\trans[i]$ does not exhibit concurrent writes in $\schedule$.
    \end{itemize}
\end{definition}

We then say that the schedule $s$ is allowed under \mvrc (respectively, \si) if every transaction is allowed under \mvrc (respectively, \si) in $s$. The latter definitions correspond to the ones in the literature (see, e.g., \cite{DBLP:conf/pods/Fekete05,vldbpaper}).
    {We emphasize that our definition of \rc is based on concrete implementations over multiversion databases, found in e.g. PostgreSQL, and should therefore not be confused with different interpretations of the term Read Committed, such as lock-based implementations~\cite{DBLP:conf/sigmod/BerensonBGMOO95} or more abstract specifications covering a wider range of concrete implementations (see, e.g.,~\cite{DBLP:conf/icde/AdyaLO00}). In particular, abstract specifications such as~\cite{DBLP:conf/icde/AdyaLO00} do not require the read-last-committed property, thereby facilitating implementations in distributed settings, where read operations are allowed to observe outdated versions. When studying robustness, such a broad specification of \rc is not desirable, since it allows for a wide range of schedules that are not conflict-serializable.
    }

\begin{figure}
    \begin{center}
        \begin{tikzpicture}[remember picture]
            \node[anchor=west] at (0,0)
            {
                $
                    \phantom{{\R[2]{\y}}}\,\,
                    \subnode{sdnode11}{\W[1]{\z}}\,\,
                    \phantom{{\W[3]{\y}}\,\CT[3]}\,\,
                    \subnode{sdnode12}{\R[1]{\x}}\,{\CT[1]\,}\,\,
                    \phantom{{\W[2]{\x}}\,\CT[2]}\,\,
                $
            };
            \node[anchor=west] at (0,-0.5)
            {
                $
                    \subnode{sdnode21}{\R[2]{\y}}\,\,
                    \phantom{\W[1]{\z}}\,\,
                    \phantom{{\W[3]{\y}}\,\CT[3]}\,\,
                    \phantom{{\R[1]{\x}}\,{\CT[1]\,}}\,\,
                    \subnode{sdnode22}{\W[2]{\x}}\,\CT[2]\,\,
                $
            };
            \node[anchor=west] at (0,-1)
            {
                $
                    \phantom{{\R[2]{\y}}}\,\,
                    \phantom{\W[1]{\z}}\,\,
                    \subnode{sdnode31}{\W[3]{\y}}\,\CT[3]\,\,
                    \phantom{{\R[1]{\x}}\,{\CT[1]\,}}\,\,
                    \phantom{{\W[2]{\x}}\,\CT[2]}\,\,
                $
            };



            \draw[->,dashed] (sdnode12) to[out=-90,in=180] (sdnode22);
            \draw[->,dashed] (sdnode21) to[out=-90,in=180] (sdnode31);

            \node at (-0.5,0) {$\trans[1]:$};
            \node at (-0.5,-0.5) {$\trans[2]:$};
            \node at (-0.5,-1) {$\trans[3]:$};
        \end{tikzpicture}
    \end{center}
    \caption{\label{fig:ex:dangerousstructure}
        {Example of a dangerous structure $\trans[1] \rightarrow \trans[2] \rightarrow \trans[3]$ with the required rw-antidependencies represented through dashed arrows.}}
\end{figure}

\inFullVersion{
\begin{figure}
    \begin{center}
        \begin{tikzpicture}[remember picture]
            \node[anchor=west] at (0,0)
            {
                $
                    \phantom{{\R[2]{\y}}}\,\,
                    \phantom{{\W[3]{\y}}\,\CT[3]}\,\,
                    \subnode{s2dnode11}{\R[1]{\z}}\,\,
                    \subnode{s2dnode12}{\R[1]{\x}}\,{\CT[1]\,}\,\,
                    \phantom{{\W[2]{\x}}\,\CT[2]}\,\,
                $
            };
            \node[anchor=west] at (0,-0.5)
            {
                $
                    \subnode{s2dnode21}{\R[2]{\y}}\,\,
                    \phantom{{\W[3]{\y}}\,\CT[3]}\,\,
                    \phantom{\R[1]{\z}}\,\,
                    \phantom{{\R[1]{\x}}\,{\CT[1]\,}}\,\,
                    \subnode{s2dnode22}{\W[2]{\x}}\,\CT[2]\,\,
                $
            };
            \node[anchor=west] at (0,-1)
            {
                $
                    \phantom{{\R[2]{\y}}}\,\,
                    \subnode{s2dnode31}{\W[3]{\y}}\,\CT[3]\,\,
                    \phantom{\R[1]{\z}}\,\,
                    \phantom{{\R[1]{\x}}\,{\CT[1]\,}}\,\,
                    \phantom{{\W[2]{\x}}\,\CT[2]}\,\,
                $
            };



            \draw[->,dashed] (s2dnode12) to[out=-90,in=180] (s2dnode22);
            \draw[->,dashed] (s2dnode21) to[out=-90,in=180] (s2dnode31);

            \node at (-0.5,0) {$\trans[1]:$};
            \node at (-0.5,-0.5) {$\trans[2]:$};
            \node at (-0.5,-1) {$\trans[3]:$};
        \end{tikzpicture}
    \end{center}
    \caption{\label{fig:ex:dangerousstructuretwo}
        {Example of a dangerous structure $\trans[1] \rightarrow \trans[2] \rightarrow \trans[3]$ where $\trans[1]$ is a read-only transaction. The required rw-antidependencies are represented through dashed arrows.}}
\end{figure}
}

While \mvrc and \si are defined on the granularity of a single transaction,
\ssi enforces a global condition on the schedule as a whole.
For this, recall the concept of dangerous structures from~\cite{DBLP:conf/sigmod/CahillRF08}:
three transactions $\trans[1],\trans[2],\trans[3] \in \transset$ (where $\trans[1]$ and $\trans[3]$ are not necessarily different) form a \emph{dangerous structure} $\trans[1] \rightarrow \trans[2] \rightarrow \trans[3]$ in $\schedule$ if:
\begin{itemize}
    \item there is a rw-antidependency from $\trans[1]$ to $\trans[2]$ and from $\trans[2]$ to $\trans[3]$ in $\schedule$;
    \item $\trans[1]$ and $\trans[2]$ are concurrent in $\schedule$;
    \item $\trans[2]$ and $\trans[3]$ are concurrent in $\schedule$;
    \item ${\CT[3] \schord \CT[1]}$ and $\CT[3] \schords \CT[2]$; and
    \item {if $\trans[1]$ is read-only, then $\CT[3] \schords \tfirst(\trans[1])$.}
\end{itemize}

Note that this definition of dangerous structures slightly extends upon the one in~\cite{DBLP:conf/sigmod/CahillRF08}, where it is not required for $\trans[3]$ to commit before $\trans[1]$ and $\trans[2]$. In the full version~\cite{DBLP:journals/tods/CahillRF09} of that paper, it is shown that, if all transactions are allowed under \si, such a structure can only lead to non-serializable schedules if $\trans[3]$ commits first.
    {Furthermore, Ports and Grittner~\cite{PortsGrittner2012} show that if $\trans[1]$ is a read-only transaction, this structure can only lead to non-serializable behavior if $\trans[3]$ commits before $\trans[1]$ starts.}
Actual implementations of \ssi (e.g., PostgreSQL~\cite{PortsGrittner2012}) therefore include this optimization when monitoring for dangerous structures to reduce the number of aborts due to false positives.
It is interesting to note that presence of a dangerous structure on itself does not necessarily mean that the schedule $\schedule$ is non-conflict-serializable, as our definition does not require a cycle in the serialization graph $\cg{\schedule}$. However, if all transactions are allowed under \si, then every cycle in $\cg{\schedule}$ implies a dangerous structure as part of the cycle~\cite{PortsGrittner2012,DBLP:journals/tods/FeketeLOOS05}. Stated differently, the absence of dangerous structures is a sufficient condition for conflict-serializability when all transactions are allowed under \si.

\inFullVersion{
\begin{example}
        Figure~\ref{fig:ex:dangerousstructure} provides an example of a dangerous structure $\trans[1] \rightarrow \trans[2] \rightarrow \trans[3]$. Notice in particular that $\trans[1]$ is not a read-only transaction, thereby allowing $\trans[1]$ and $\trans[3]$ to be concurrent. Indeed, if we would replace the write operation $\W[1]{\z}$ by a read operation $\R[1]{\z}$ in Figure~\ref{fig:ex:dangerousstructure}, the result would no longer be a dangerous structure. An example of a dangerous structure where $\trans[1]$ is a read-only transaction is given in Figure~\ref{fig:ex:dangerousstructuretwo}.
    \hfill $\Box$
\end{example}
}

We are now ready to define when a schedule is allowed under a (mixed) allocation of isolation levels.

\begin{definition} \label{def:mixed:schedule}
    A schedule $\schedule$ over a set of transactions $\transset$ \emph{is allowed under an allocation $\alloc$} over $\transset$ if:
    \begin{itemize}
        \item for every transaction $\trans[i] \in \transset$ with $\alloc(\trans[i]) = \mvrc$, $\trans[i]$ is allowed under \mvrc;
        \item for every transaction $\trans[i] \in \transset$ with $\alloc(\trans[i]) \in \{\si,\ssi\}$, $\trans[i]$ is allowed under \si; and
        \item there is no dangerous structure $\trans[i] \rightarrow \trans[j] \rightarrow \trans[k]$ in $\schedule$ formed by three (not necessarily different) transactions $\trans[i], \trans[j], \trans[k] \in \{\trans[] \in \transset \mid \alloc(\trans[]) = \ssi\}$.
    \end{itemize}
\end{definition}

We denote the allocation mapping all transactions to \mvrc (respectively, \si) by $\allocrc$ (respectively, $\allocsi$).
We illustrate some of the just introduced notions through an example.
\begin{example}\label{ex:schedule}

    Consider the schedule $s$ in Figure~\ref{fig:ex:schedule}.
    Transaction $T_1$ is concurrent with $T_2$ and $T_4$, but not with
    $T_3$; all other transactions are pairwise concurrent with each other.
    The second read operation of $T_4$ is a read-last-committed relative to
    itself but not relative to the start of $T_4$.
        {The read operation $\R[2]{\y}$ of $\trans[2]$ is read-last-committed relative to the start of $\trans[2]$, but not relative to itself, so an allocation mapping $\trans[2]$ to \rc is not allowed.}
    All other read operations
        {are read-last-committed relative to both themselves and the start of the corresponding transaction.}
    None of the transactions exhibits a dirty write.
    Only transaction $T_4$ exhibits a concurrent write (witnessed by the write
    operation $\UP[2]{\x}$ in $T_2$). Due to this, an allocation mapping $T_4$ on
    \si or \ssi is not allowed.
    The transactions $T1\to T2 \to T3$ form a dangerous structure, therefore
    an allocation mapping all three transactions $T_1,T_2,T_3$ on \ssi is not
    allowed.
    All other allocations, that is, mapping $T_4$ on \rc, {$\trans[2]$ on \si or \ssi} and at least one of
    $T_1,T_2,T_3$ on \rc or \si, is allowed. \hfill $\Box$

    %
    %
    %
    %
    %
    %
    %

\end{example}

\subsection{\PTranss{}}
\label{sec:templates}

\Ptranss{} are transactions where operations are defined over typed variables.
Types of variables are relation names in $\schRel$ and indicate that variables can only be instantiated by tuples from the respective type.
We fix an infinite set of variables $\variables$ that is disjoint from $\objects$. Every variable $\vx\in\variables$ has an associated relation name in $\schRel$ as type that we denote by $\type{\vx}$.

\begin{definition}\label{def:template}
A \emph{\ptrans{}} $\tau$ is a transaction over $\variables$. 
{In addition, for every operation $o$ in $\tau$ over a variable $\vx$,
$\ReadSet{o}\subseteq \Attr{\type{\vx}}$ and $\WriteSet{o}\subseteq \Attr{\type{\vx}}$}.
\end{definition}
For an operation $o$ in a \ptrans{} $\templ[]$, we denote by $\myvar{o}$ the variable over which $o$ is defined.
Notice that operations in \ptranss{} are defined over typed variables whereas they are over $\objects$ in transactions.
Indeed, the \ptrans{} for Balance in Figure~\ref{fig:smallbank-abstract-syntax} contains a read operation $o=\R[]{\vx: \Account\{N,C\}}$.
As explained in Section~\ref{sec:SBexample}, the notation $\mathtt{\vx}: \Account\{N,C\}$ is a shorthand for $\type{\mathtt{\vx}}=\Account$ and $\ReadSet{o}=\{N,C\}$.

Recall that we denote variables by capital letters $\mathtt{\vx},\mathtt{\vy},
\mathtt{\vz}$ and tuples by small letters $\x,\y$. 
A variable assignment $\mu$ is a mapping from $\variables$ to $\objects$
such that $\mu(\vx)\in \objects_{\type{\vx}}$.
By $\tmap (\templ[])$, we denote the transaction $\trans[]$ obtained by replacing each variable $\vx$ in $\templ[]$ with $\tmap (\vx)$. A variable assignment for a database $\db$ maps every variable to a tuple occurring in a relation in $\db$.
We say that a transaction $\trans[]$ is \emph{instantiated} from a template $\templ[]$ over a database $\db$ if there is a variable assignment $\mu$ for $\db$ such that $\trans[] = \mu(\templ[])$.
As a slight abuse of notation, we will frequently write $\mu(o)$ for an operation $o$ in $\templ[]$ to denote the corresponding operation in $\trans[]$.

A set of transactions $\transset$ is \emph{consistent} with a set of \ptranss{} $\templset$ {and 
database $\db$}, if every transaction in $\transset$ is instantiated from a template in $\templset$ over $\db$. That is, for every transaction $\trans[]$ in $\transset$ there is a \ptrans{} $\tau\in \templset$ and a variable assignment $\mu_{\trans[]}$ 
for $\db$ such that $\mu_{\trans[]}(\tau) = \trans[]$.

We extend the notion of allocations towards transaction templates. For a class of isolation levels $\isolationlevel$,
a \emph{template $\isolationlevel$-allocation $\alloc^\templset$} for a set of transaction templates $\templset$ is a function mapping each template $\templ[] \in \templset$ onto an isolation level $\alloc^\templset(\templ[]) \in \isolationlevel$. When $\isolationlevel$ is not important or clear from the context, we will frequently refer to $\alloc^\templset$ as a template allocation rather than template $\isolationlevel$-allocation.

Let $\transset$ be a set of transactions consistent with a set of transaction templates $\templset$ and a database $\db$, and let $\alloc^\templset$ be a template $\isolationlevel$-allocation for $\templset$. 
An allocation $\alloc$ for $\transset$ is \emph{consistent} with $\alloc^\templset$ and $\db$ if for every transaction $\trans[] \in \transset$ there is a template $\templ[] \in \templset$ such that $\trans[]$ is instantiated from $\templ[]$ over $\db$ and $\alloc(\trans[]) = \alloc^\templset(\templ[])$.

\subsection{Transaction and Template Robustness}
\label{sec:def-robustness}

We first define the robustness property~\cite{DBLP:conf/concur/0002G16} (also called \emph{acceptability} in~\cite{DBLP:conf/pods/Fekete05,DBLP:journals/tods/FeketeLOOS05}) over a given set of transactions $\transset$, which guarantees serializability for all schedules over $\transset$ for a given allocation.

\begin{definition}[Transaction robustness]
    \label{def:robustness}
    A set of transactions $\transset$ is \emph{robust} against an allocation $\alloc$ for $\transset$
    if every schedule for\/ $\transset$ that is allowed under $\alloc$ is
    conflict-serializable.
\end{definition}
We refer to $\alloc$ as a \emph{robust allocation}.
The \emph{(transaction) robustness problem} is then to decide whether a given allocation for a set of transactions $\transset$ is a robust allocation. A polynomial time algorithm for transaction robustness is given in \cite{DBLP:conf/pods/VandevoortKN23}. 

We next lift robustness to the level of templates by requiring transaction robustness for all possible template instantiations and all possible databases. 
Let $\workload$ be a set of \ptranss{} and $\db$ be a database.
Then, $\workload$ is \emph{robust against a template allocation $\alloc^\templset$ over $\db$} if for every set of transactions $\transset$ that is consistent with $\workload$ and $\db$ and for every allocation $\alloc$ for $\transset$ consistent with $\alloc^\templset$ and $\db$, it holds that $\transset$ is robust against $\alloc$.

\begin{definition}[Template robustness]\label{def:template_robustness}
 A set of \ptranss{} 
$\templset$ is \emph{robust} against a template allocation $\alloc^\templset$ for $\templset$ if\/ $\templset$ is robust against $\alloc^\templset$ for every database $\db$.
\end{definition}

\inFullVersion{
\new{

    \begin{figure*}

        \begin{center}

            \begin{tikzpicture}[remember picture]
                \node[anchor=west](nodeT0) at (0,0.5) {$\sstart$};
                \node[anchor=west] at (0,0)
                {$\phantom{\sstart\,} \subnode{nodeT11}{\R[1]{\x_4}}\, \subnode{nodeT12}{\R[1]{\y_1}}\,
                \phantom{\R[2]{\x_3}\, \UP[2]{\y_1}\, \CT[2]
                \R[3]{\x_3}\, \R[3]{\y_1}\, \R[3]{\z_2}\, \UP[3]{\z_2}\, \CT[3]}
                \subnode{nodeT13}{\R[1]{\z_2}}\, \CT[1]$};


                \node[anchor=west] at (0,-0.5)
                {$\phantom{\sstart\, \R[1]{\x_4}\, \R[1]{\y_1}}
                \subnode{nodeT21}{\R[2]{\x_3}}\, \subnode{nodeT22}{\UP[2]{\y_1}}\, \CT[2]$};

                \node[anchor=west] at (0,-1)
                {$\phantom{\sstart\, \R[1]{\x_4}\, \R[1]{\y_1}
                \R[2]{\x_3}\, \UP[2]{\y_1}\, \CT[2]}
                \subnode{nodeT31}{\R[3]{\x_3}}\, \subnode{nodeT32}{\R[3]{\y_1}}\, \subnode{nodeT33}{\R[3]{\z_2}}\, \subnode{nodeT34}{\UP[3]{\z_2}}\, \CT[3]$};
                \draw[->,solid,out=90,in=90] (nodeT32) to (nodeT22);
                \draw[->,solid,out=-90,in=90] (nodeT13) to (nodeT34);
                
                \draw[->,-{Triangle[open,length=1.5mm,width=1.5mm]},solid,double,in=-120,out=-100] (nodeT0) to (nodeT22);
                \draw[->,-{Triangle[open,length=1.5mm,width=1.5mm]},solid,double,in=130,out=10] (nodeT0) to (nodeT34);
                \node at (-0.5,0) {$\trans[1]:$};
                \node at (-0.5,-0.5) {$\trans[2]:$};
                \node at (-0.5,-1) {$\trans[3]:$};
            \end{tikzpicture}

        \end{center}
        \caption{\label{fig:ex:sched:smallbank}
            \new{A schedule $\schedule$ with $\schvf$ (single lines) and $\schvord$ (double lines) represented through
            arrows. Single line arrows towards $\sstart$ are omitted for clarity. Transaction $\trans[1]$ is instantiated from the \ptrans{} {Balance}, $\trans[2]$ is instantiated from {TransactSavings}, and $\trans[3]$ is instantiated from {WriteCheck}.}
        }

    \end{figure*}

    \begin{example}
        \label{ex:sb:robust}
        Consider the Template allocation $\alloc^\templset$ for the SmallBank \ptranss{} in Figure~\ref{fig:smallbank-abstract-syntax} where $\alloc^\templset(\Balance) = \mvrc$, and all other templates are mapped to \si.
        Then, the set of three transactions $\transset = \{\trans[1],\trans[2],\trans[3]\}$ in Figure~\ref{fig:ex:sched:smallbank} is consistent with the SmallBank \ptranss{}. Furthermore, the allocation $\alloc$ for $\transset = \{\trans[1],\trans[2],\trans[3]\}$ with $\alloc(\trans[1]) = \mvrc$, $\alloc(\trans[2]) = \si$ and $\alloc(\trans[3]) = \si$ is consistent with $\alloc^\templset$. It now follows that the SmallBank \ptranss{} are not robust against $\alloc^\templset$. Indeed, Schedule $\schedule$ in Figure~\ref{fig:ex:sched:smallbank} is allowed under $\alloc$ and not conflict-serializable, thereby witnessing non-robustness.
    \end{example}
}    
}

\subsection{Deciding Template Robustness}
\label{sec:robustness}


We are now ready to present our first algorithmic result: a polynomial-time algorithm for template robustness. In   Section~\ref{sec:allocation}, we demonstrate how this algorithm can be applied to identify the lowest robust allocation.

\smallskip
\noindent
{\bf Outline of approach.}
We recall that template robustness is defined over all possible database instances. Consequently, any approach that considers all possible instantiations of transaction templates and then applies the transaction robustness algorithm from \cite{DBLP:conf/pods/VandevoortKN23} is infeasible due to the infinite number of possible instantiations. The algorithm proposed in this paper addresses this challenge using the following approach.  We first introduce the concept of a sequence of potentially conflicting quadruples (which can be seen as an abstraction of a path in a serialization graph induced by the templates). We then obtain some conditions that characterize when this sequence can be converted to a counterexample schedule, that is, a
non-conflict-serializable schedule that is still allowed under the given allocation. Furthermore, the schedule is of a very specific form to which we refer as a split schedule and only requires the existence of four different tuples per relation in the database. For the decision problem it then suffices to check for the existence of a sequence satisfying the above mentioned conditions, for which we present a polynomial time algorithm.


\smallskip
\noindent
{\bf Sequence of potentially conflicting quadruples.} 
Let $\templ[i]$ and $\templ[j]$ be two (not necessarily different) templates in $\templset$, and let $o_i$ and $p_j$ be two operations in $\templ[i]$ and $\templ[j]$, respectively.
We then say that $o_i$ is \emph{potentially conflicting} with $p_j$ if $o_i$ and $p_j$ are over variables of the same type (i.e., $\type{\myvar{o_i}} = \type{\myvar{p_j}}$) and at least one of the following conditions holds:
\begin{itemize}
    \item (potential ww-conflict) $\WriteSet{o_i} \cap \WriteSet{p_j} \neq \emptyset$;
    \item (potential wr-conflict) $\WriteSet{o_i} \cap \ReadSet{p_j} \neq \emptyset$; or
    \item (potential rw-conflict) $\ReadSet{o_i} \cap \WriteSet{p_j} \neq \emptyset$.
\end{itemize}
In this case, we also say that the tuple $(\templ[i], o_i, p_j, \templ[j])$ is a \emph{potentially conflicting quadruple} over $\templset$.
Intuitively, a potentially conflicting quadruple represents a pair of operations that leads to conflicting operations whenever the corresponding variables are instantiated with the same tuple by a variable assignment. We will sometimes refer to the operation $o$ as an \emph{outgoing operation} and the operation $p$ as an \emph{incoming operation} in the quadruple $(\templ[i], o, p, \templ[j])$.

Towards our algorithm, we consider \emph{sequences of potentially conflicting quadruples}
$\seqpcq = (\templ[1], o_1, p_2, \templ[2]), (\templ[2], o_2, p_3, \templ[3]), \ldots,\allowbreak (\templ[n-1], o_{n-1},\allowbreak p_n, \templ[n]), (\templ[n], o_n, p_1, \templ[1])$
over a set of templates $\templset$, where each tuple is a potentially conflicting quadruple over $\templset$, and where multiple occurrences of the same template in $\seqpcq$ are allowed (i.e., $\templ[i] = \templ[j]$ is allowed, even when $i \neq j$). Notice in particular that $\seqpcq$ starts and ends with the same template $\templ[1]$.

\inFullVersion{
\new{
    \begin{example}
        \label{ex:sb:seq}
        A possible sequence of potentially conflicting quadruples over the SmallBank transaction templates given in Figure~\ref{fig:smallbank-abstract-syntax} is
        \begin{multline*}
            \seqpcq_1 = (\templ[1], \R[]{\vy: \Savings\ListAttr{C, B}}, \UP[]{\vy: \Savings\ListAttr{C, B}\ListAttr{B}}, \templ[2]),\\
            (\templ[2], \UP[]{\vy: \Savings\ListAttr{C, B}\ListAttr{B}}, \R[]{\vy: \Savings\ListAttr{C, B}}, \templ[3]),\\
            (\templ[3], \UP[]{\vz: \Checking\ListAttr{C, B}\ListAttr{B}}, \R[]{\vz: \Checking\ListAttr{C, B}}, \templ[1]),
        \end{multline*}
        with $\templ[1] = \text{Balance}$, $\templ[2] = \text{TransactSavings}$ and $\templ[3] = \text{WriteCheck}$.
        In this sequence, the first quadruple is based on a potential rw-conflict, whereas the other two quadruples are based on potential wr-conflicts.
    \end{example}
}
}

We will use a sequence $\seqpcq$ of potentially conflicting quadruples to construct a non-conflict-serializable schedule, where a cycle in the serialization graph is formed by the operations in $\seqpcq$. For this to work, care must be taken to ensure that the variables of the operations occurring in a potentially conflicting quadruple are assigned the same tuple in the database instance. Indeed, otherwise there would be no dependency between these operations in the schedule. Let $o$ and $p$ be two operations from templates $\templ[i]$ and $\templ[j]$ respectively, where $\templ[i]$ and $\templ[j]$ both occur in a sequence of potentially conflicting quadruples $\seqpcq = (\templ[1], o_1, p_2, \templ[2]), \ldots, (\templ[n], o_n, p_1, \templ[1])$. We then say that the variables $\myvar{o}$ and $\myvar{p}$ are \emph{connected} in $\seqpcq$ if
\begin{itemize}
    \item $i = j$ and $\myvar{o} = \myvar{p}$ \emph{(connected within the same template)};
    \item there exists a quadruple $(\templ[i], o, p, \templ[j])$ or $(\templ[j], p, o, \templ[i])$ in $\seqpcq$ \emph{(connected between templates)}; or
    \item there exists a variable $\vx$ occurring in a template in $\seqpcq$ such that both $\myvar{o}$ and $\myvar{p}$ are connected to $\vx$ \emph{(transitivity)}.
\end{itemize}
Intuitively, connected variables must be assigned the same tuple for the variable assignments to be valid while ensuring that the desired dependencies are in place.

\smallskip
\noindent
{\bf Mapping to a database with four elements per relation.}
Before we define variable assignments $\mu_i$ for each template $\templ[i]$ in $\seqpcq$, we first introduce a special database instance $\db_4$ over the considered schema $\schRel$ containing four tuples per relation in $\schRel$. We refer to these tuples as $\x^R_1$, $\x^R_2$, $\x^R_3$, and $\x^R_4$ for each $R \in \schRel$ and use $\db_4$ to construct the variable assignments $\mu_i$. 
To this end, we first define four \emph{type mappings} $c_1$, $c_2$, $c_3$ and $c_4$ that map each relation $R \in \schRel$ to a tuple of the corresponding type in $\db_4$. Formally, we set $c_i(R) = \x^R_i$ for each $i \in \{1,2,3,4\}$ and $R \in \schRel$.
For each template $\templ[i]$ in $\seqpcq$, we define 
the \emph{canonical variable assignments} $\mu_i$ over $\db_4$ as follows: 
\[
    \mu_1(\vx) = \left\{
        \begin{array}{ll}
            c_1(\type{\vx}) & \text{if $\vx$ is connected to $\myvar{o_1}$,} \\
            c_2(\type{\vx}) & \text{if $\vx$ is connected to $\myvar{p_1}$ and}\\
            &\text{not to $\myvar{o_1}$,} \\ 
            c_4(\type{\vx}) & \text{otherwise.} 
        \end{array}\right.
\]
For every $\mu_i$ with $1<i\leq m$,
\[
    \mu_i(\vx) = \left\{
        \begin{array}{ll}
            c_1(\type{\vx}) & \text{if $\vx$ is connected to $\myvar{o_1}$}, \\
            c_2(\type{\vx}) & \text{if $\vx$ is connected to $\myvar{p_1}$}\\
            &\text{not to $\myvar{o_1}$,} \\ 
            c_3(\type{\vx}) & \text{otherwise.} 
        \end{array}\right.
\]
By construction, type mapping $c_4$ is used exclusively for $\templ[1]$, whereas $c_3$ is only used for $\templ[2], \ldots, \templ[n]$;
$c_1$ is used for all variables connected to $\myvar{o_1}$; and, $c_2$ is used for all variables connected to $\myvar{p_1}$, unless $\myvar{o_1}$ and $\myvar{p_1}$ are connected in $\seqpcq$ in which case
$c_1$ is used as well as by transitivity variables connected to $\myvar{p_1}$ are also connected to $\myvar{o_1}$. 

\smallskip
\noindent
{\bf From a sequence to a split schedule.}
For a given sequence $\seqpcq$ of potentially conflicting quadruples over a set of templates $\templset$, we define the \emph{canonical set of transactions $\transset_\seqpcq$} as the set obtained by applying the canonical variable assignment $\mu_i$ to each template $\templ[i]$ occurring in $\seqpcq$.
A template allocation $\alloc^\templset$ over $\templset$ then induces a \emph{canonical allocation $\alloc_\seqpcq$} over $\transset_\seqpcq$ in a natural way: for each template $\templ[i]$ in $\seqpcq$, we allocate the corresponding transaction $\trans[i] = \mu_i(\templ[i])$ to the isolation level $\alloc_\seqpcq(\trans[i]) = \alloc^\templset(\templ[i])$. By construction, $\transset_\seqpcq$ is consistent with $\templset$ and $\db_4$, and $\alloc_\seqpcq$ is consistent with $\alloc^\templset$ and $\db_4$.
{For a transaction $\trans[]$ and operation $o \in \trans[]$, we denote by $\prefix{\trans[]}{o}$ the subsequence of $\trans[]$ containing all operations up to and including $o$, and by $\postfix{\trans[]}{o}$ the subsequence of $\trans[]$ containing all operations strictly after $o$. This notation extends to templates in a natural way.}

\begin{definition}\label{def:templ_split_sched}
    Let $\templset$ be a set of transaction templates, $\alloc^\templset$ a template allocation over $\templset$, and
    $\seqpcq = (\templ[1], o_1, p_2, \templ[2]), \ldots, (\templ[n], o_n, p_1, \templ[1])$ a sequence of potentially conflicting quadruples over $\templset$. A \emph{template split schedule $\schedule$ for $\templset$ and $\alloc^\templset$ induced by $\seqpcq$} is a multiversion schedule over the canonical set of transactions $\transset_\seqpcq$ of the 
    form:
    \[\prefix{\mu_1(\templ[1])}{\mu_1(o_1)} \cdot \mu_2(\templ[2]) \cdot \ldots \cdot \mu_n(\templ[n]) \cdot \postfix{\mu_1(\templ[1])}{\mu_1(o_1)},\]
    where
    \begin{enumerate}
        \item $\schedule$ is allowed under the canonical allocation $\alloc_\seqpcq$ induced by 
        $\alloc^\templset$;
        \item $\dependson[\schedule]{\mu_i(o_i)}{\mu_j(p_j)}$ for each quadruple $(\templ[i], o_i, p_j, \templ[j]) \in \seqpcq$; and
        \item there is no operation in $\mu_1(\templ[1])$ conflicting with an operation in any of the transactions $\mu_3(\templ[3]), \ldots, \mu_{n-1}(\templ[n-1])$.
    \end{enumerate}

\end{definition}

\inFullVersion{
\new{
    \begin{example}
        \label{ex:sb:splitsched}
        Continuing our running example based on the SmallBank transaction templates given in Figure~\ref{fig:smallbank-abstract-syntax},
        the special database instance $\db_4$ contains four tuples per relation, which we denote as $\x_i$, $\y_i$ and $\z_i$ with $i \in \{1,2,3,4\}$ for the relations \Account, \Savings and \Checking, respectively.
        The canonical set of transactions for the sequence of potentially conflicting quadruples $\seqpcq_1$ from Example~\ref{ex:sb:seq} then corresponds to the three transactions introduced in Example~\ref{ex:sb:robust}.
        Figure~\ref{fig:ex:sched:smallbank} furthermore shows the template split schedule $\schedule$ for the SmallBank templates and template allocation $\alloc^\templset$ from Example~\ref{ex:sb:robust} induced by $\seqpcq_1$.
        Notice in particular that the schedule is allowed under the canonical allocation (i.e., $\trans[1]$ is allowed under \mvrc, and $\trans[2]$ and $\trans[3]$ are allowed under \si), and that the each quadruple in $\seqpcq_1$ corresponds to a dependency in $\schedule$.
    \end{example}
}
}

Notice that the cycle of dependencies between the operations in $\seqpcq$ implies that such a schedule is not conflict-serializable. 
The next proposition then readily follows:
\begin{proposition}\label{prop:split:non-conflict-serializable}
    Let $\schedule$ be a template split schedule for a set of templates $\templset$ and template allocation $\alloc^\templset$ over $\templset$ induced by a sequence of potentially conflicting quadruples $\seqpcq$. Then, $\schedule$ is non-conflict-serializable and allowed under $\alloc^\templset$.
\end{proposition}

\smallskip
\noindent
{\bf Conditions characterizing the existence of a counterexample schedule.}
We introduce a set of conditions that must be satisfied by a sequence of potentially conflicting quadruples for a template split schedule to exist. 
\begin{propositionrep}
    \label{prop:conds:split}
    Let $\templset$ be a set of transaction templates, $\alloc^\templset$ a template allocation over $\templset$, and
    $\seqpcq = (\templ[1], o_1, p_2, \templ[2]), \ldots, (\templ[n], o_n, p_1, \templ[1])$ a sequence of potentially conflicting quadruples over $\templset$. A template split schedule for $\templset$ and $\alloc^\templset$ induced by $\seqpcq$ exists if and only if the following conditions hold:
    \begin{enumerate}
        \item \label{c:1} there is no operation $o$ in $\templ[1]$ potentially conflicting with an operation $p$ in any of the templates $\templ[3],\ldots, \templ[n-1]$ with $\myvar{o}$ and $\myvar{p}$ connected in $\seqpcq$;
        \item \label{c:2} there is no write operation $o$ in $\prefix{\templ[1]}{o_1}$ potentially ww-conflicting with a write operation $p$ in $\templ[2]$ or $\templ[n]$ where $\myvar{o}$ and $\myvar{p}$ are connected in $\seqpcq$;
        \item \label{c:3} if $\alloc^\templset(\templ[1]) \in \{\si, \ssi\}$, then there is no write operation $o$ in $\postfix{\templ[1]}{o_1}$ potentially ww-conflicting with a write operation $p$ in $\templ[2]$ or $\templ[n]$ with $\myvar{o}$ and $\myvar{p}$ connected in $\seqpcq$; 
        \item \label{c:4} $o_{1}$ is potentially rw-conflicting with $p_{2}$;
        \item \label{c:5} $o_{n}$ is potentially rw-conflicting with $p_{1}$ or ($\alloc^\templset(\templ[1]) = \rc$ and $o_1 <_{\templ[1]} p_1$);
        \item \label{c:6} $\alloc^\templset(\templ[1]) \neq \ssi$ or $\alloc^\templset(\templ[2]) \neq \ssi$ or $\alloc^\templset(\templ[n]) \neq \ssi$;
        \item \label{c:7} if $\alloc^\templset(\templ[1]) = \ssi$ and $\alloc^\templset(\templ[2]) = \ssi$, then there is no operation $o$ in $\templ[1]$ potentially wr-conflicting with an operation $p$ in $\templ[2]$ with $\myvar{o}$ and $\myvar{p}$ connected in $\seqpcq$; and
        \item \label{c:8} if $\alloc^\templset(\templ[1]) = \ssi$ and $\alloc^\templset(\templ[n]) = \ssi$, then there is no operation $o$ in $\templ[1]$ potentially rw-conflicting with an operation $p$ in $\templ[n]$ with $\myvar{o}$ and $\myvar{p}$ connected in $\seqpcq$.
    \end{enumerate}
\end{propositionrep}

\begin{proof}
    \emph{(if)} Assuming the conditions hold, we construct a schedule $\schedule = (\schop, \schord, \schvord, \schvf)$ and argue that it is a valid template split schedule. Notice that $\schop$ and $\schord$ are fixed by $\seqpcq$ and the definition of template split schedule induced by $\seqpcq$ (cf.\ Definition~\ref{def:templ_split_sched}).
    The remaining $\schvord$ and $\schvf$ can be derived from the canonical allocation $\alloc_\seqpcq$ over $\transset_\seqpcq$. Indeed, by Definition~\ref{def:templ_split_sched}, our intention is to construct a schedule $\schedule$ such that $\schedule$ is allowed under $\alloc_\seqpcq$, implying that $\schvord$ and $\schvf$ must adhere to the corresponding isolation levels. In particular, \emph{(i)} the version order $\schvord$ should be chosen such that each write operation in $\schedule$ respects the commit order, and \emph{(ii)} the version function $\schvf$ should be constructed such that the read operations are read-last-committed relative to the read operation itself (in case of RC), or relative to the start of the transaction (in case of SI or SSI). It is straightforward to verify that these two requirements always lead to a single unique definition of $\schvord$ and $\schvf$, although it should be noted that this construction on itself does not yet guarantee $\schedule$ to be allowed under $\alloc_\seqpcq$.
    The remainder of the argument relies on showing that the constructed schedule is indeed a template split schedule due to the assumed conditions. We prove each of the three conditions stated in Definition~\ref{def:templ_split_sched} below.
    
    \medskip\noindent
    \underline{\textit{$\schedule$ is allowed under $\alloc_\seqpcq$.}} Let $\trans[i] = \mu_i(\templ[i])$ for some $i \in \{1, \ldots, n\}$ be a transaction in $\schedule$. If $\alloc_\seqpcq(\trans[i]) = \rc$. The fact that each write operation in $\trans[i]$ respects the commit order of $\schedule$ is immediate by our choice of $\schvord$. Similarly, our choice of $\schvf$ ensures that each read operation $b_i$ in $\trans[i]$ is read-last-committed in $\schedule$ relative to $b_i$ itself, and hence we only need to argue that $\trans[i]$ does not exhibit a dirty write in $\schedule$.
    From the construction of $\schord[]$, such a dirty write can only occur if $i \neq 1$.
    Towards a contradiction, assume that $\trans[i]$ exhibits a dirty write in $\schedule$. From the construction of $\schord[]$, there is a write operation $b_1 \in \prefix{\trans[1]}{\mu_1(o_1)}$ conflicting with a write operation $a_i \in \trans[i]$. By construction of $\mu_1$ and $\mu_i$, these operations can only be conflicting if the corresponding variables in $\tau_1$ and $\tau_i$ are connected in $\seqpcq$. Condition~(\ref{c:1}) hence implies that $i \not\in \{3, \ldots, n-1\}$, whereas Condition~(\ref{c:2}) implies that $i \not\in \{2, n\}$, leading to the desired contradiction. We therefore conclude that $\trans[i]$ is indeed allowed under \rc.
    
    Next, consider the case where $\alloc_\seqpcq(\trans[i]) = \si$ or $\alloc_\seqpcq(\trans[i]) = \ssi$. Our choice of $\schvord[]$ and $\schvf[]$ again implies that the write operations in $\trans[i]$ respect the commit order and that the read operations are read-last-committed relative to the start of the transaction. We argue that $\trans[i]$ does not exhibit a concurrent write in $\schedule$. Towards a contradiction, assume there are two conflicting write operation $b_j \in \trans[j]$ and $a_i \in \trans[i]$ for some $j \neq i$ such that $b_j \schords a_i$ and $\tfirst(\trans[i]) \schords \CT[j]$. If this concurrent write is a dirty write as well, the same argument as above applies. Otherwise, if $\CT[j] \schords a_i$, such a concurrent write is only possible if $i = 1$, and in particular $a_i \in \postfix{\trans[1]}{\mu_1(o_1)}$. Similar to the previous case, we can use Conditions~(\ref{c:1}) and (\ref{c:3}) to derive a contradiction. We therefore conclude that $\trans[i]$ is indeed allowed under \si if $\alloc_\seqpcq(\trans[i]) = \si$ or $\alloc_\seqpcq(\trans[i]) = \ssi$.
    
    Lastly, we must argue that there is no dangerous structure $\trans[i] \rightarrow \trans[j] \rightarrow \trans [k]$ in $\schedule$ with  $\alloc_\seqpcq(\trans[i]) = \alloc_\seqpcq(\trans[j]) = \alloc_\seqpcq(\trans[k]) = \ssi$, since dangerous structures require $\trans[i]$ and $\trans[j]$ as well as $\trans[j]$ and $\trans[k]$ to be concurrent, this is only possible in $\schedule$ if $j = 1$ and $i,k \in \{2,n\}$. Recall from the construction of $\mu_1$, $\mu_2$ and $\mu_n$ that, because of the rw-antidependencies forming the dangerous structure, the corresponding variables must be connected in $\seqpcq$ as otherwise these operations would not be conflicting in $\schedule$. It now follows from Conditions~(\ref{c:6}), (\ref{c:7}) and (\ref{c:8}) that such a dangerous structure cannot exist in $\schedule$.

    \medskip\noindent
    \underline{\textit{$\dependson[\schedule]{\mu_i(o_i)}{\mu_j(p_j)}$ for each quadruple $(\templ[i], o_i, p_j, \templ[j]) \in \seqpcq$.}} By construction of $\mu_i$ and $\mu_j$, the operations $\mu_i(o_i)$ and $\mu_j(p_j)$ are conflicting in $\schedule$. It remains to show that $\mu_j(p_j)$ depends on $\mu_i(o_i)$ (i.e., the dependency between the two conflicting operations is in the correct direction). For dependencies $\dependson[\schedule]{\mu_1(o_1)}{\mu_2(p_2)}$ and $\dependson[\schedule]{\mu_n(o_n)}{\mu_1(p_1)}$ this follows from Condition~(\ref{c:4}) and Condition~(\ref{c:5}), respectively. For all other dependencies, the result is immediate by our construction of $\schedule$ since the two involved transactions are not concurrent in $\schedule$.

    \medskip\noindent
    \underline{\textit{There is no operation in $\mu_1(\templ[1])$ conflicting with an operation in any of the transactions $\mu_3(\templ[3]), \ldots, \mu_{n-1}(\templ[n-1])$.}} Assume towards a contradiction that there exists such an operation $b_1$ in $\mu_1(\templ[1])$ conflicting with an operation $a_j$ in some transaction $\mu_j(\templ[j])$ with $j \in \{3, \ldots, n-1\}$, and let $o$ and $p$ be the two corresponding potentially conflicting operations in $\tau_1$ and $\tau_j$, respectively. By construction of $\mu_1$ and $\mu_j$, the variables of $o$ and $p$ are connected in $\seqpcq$, as otherwise $\mu_1(\myvar{o}) \neq \mu_j(\myvar{p})$. The desired contradiction now follows from Condition~(\ref{c:1}), stating that such a pair of operations $o$ and $p$ cannot exist.

    \bigskip\noindent
    \emph{(only if)} Let $\schedule = (\schop, \schord, \schvord, \schvf)$ be the template split schedule. We show that Conditions~(\ref{c:1}) - (\ref{c:8}) hold.
    
    For~(\ref{c:1}), the result is immediate from Definition~\ref{def:templ_split_sched}(3). Indeed since $\myvar{o}$ and $\myvar{p}$ are connected, they would lead to conflicting operations in $\schedule$ by construction of the canonical variable assignments $\mu_i$.

    For~(\ref{c:2}), notice that the existence of such a pair of write operations would lead to a dirty write in $\schedule$, thereby contradicting Definition~\ref{def:templ_split_sched}(1).
    
    For~(\ref{c:3}), the result follows from the fact that the existence of such a pair of write operations would imply $\mu_1(\tau_1)$ exhibiting a concurrent write in in $\schedule$, thereby contradicting Definition~\ref{def:templ_split_sched}(1).

    For~(\ref{c:4}), we argue that $o_1$ cannot be ww-conflicting or wr-conflicting with $p_2$. The former already follows from our argument for~(\ref{c:2}), whereas the latter follows from the fact that if $o_1$ is wr-conflicting with $p_2$, then there would be a wr-dependency $\dependson[\schedule]{\mu_2(p_2)}{\mu_1(o_1)}$ instead of the rw-antidependency $\dependson[\schedule]{\mu_1(o_1)}{\mu_2(p_2)}$ implied by Definition~\ref{def:templ_split_sched}(2).

    For~(\ref{c:5}), consider the case where $o_n$ is ww-conflicting or wr-conflicting with $p_1$. The former case implies either a dirty write in $\mu_n(\templ[n])$ if $p_1 \leq_{\templ[1]} o_1$, or a concurrent write in $\mu_1(\tau_1)$ otherwise. We there conclude that this case only satisfies Definition~\ref{def:templ_split_sched}(1) when $\alloc^\templset(\tau_1) = \rc$ and $o_1 <{\templ[1]} p_1$. The latter case only implies the wr-dependency $\dependson[\schedule]{\mu_n(o_n)}{\mu_1(p_1)}$ required by Definition~\ref{def:templ_split_sched}(2) if $\alloc^\templset(\tau_1) = \rc$ and $o_1 <{\templ[1]} p_1$, as otherwise this would be a rw-antidependency in the opposite direction.

    For~(\ref{c:6}), assume towards a contradiction that $\alloc^\templset(\templ[1]) = \ssi$, $\alloc^\templset(\templ[2]) = \ssi$ and $\alloc^\templset(\templ[n]) = \ssi$. From (\ref{c:4}) and (\ref{c:5}) above, we conclude that a dangerous structure over the corresponding three transactions is formed in $\schedule$, thereby contradicting Definition~\ref{def:templ_split_sched}(1).

    For~(\ref{c:7}), consider the case where $\alloc^\templset(\templ[1]) = \ssi$ and $\alloc^\templset(\templ[2]) = \ssi$. The existence of a wr-conflict between $\mu_1(o)$ and $\mu_2(p)$ would imply a rw-antidependency from $\mu_2(\templ[2])$ to $\mu_1(\templ[1])$, thereby resulting in a dangerous structure, contradicting Definition~\ref{def:templ_split_sched}(1).

    For~(\ref{c:8}), the argument is analogous to the previous case. In particular, the existence of a rw-antidependency from $\mu_1(\templ[1])$ to $\mu_n(\templ[n])$ results in dangerous structure in $\schedule$, and is therefore not allowed by Definition~\ref{def:templ_split_sched}(1).
\end{proof}

\new{Intuitively, these conditions enforce the desired cycle of dependencies, while ensuring that the schedule is allowed under the allocation (cf.\ Definition~\ref{def:templ_split_sched}). For example, conditions (\ref{c:1}) and (\ref{c:2}) ensure that no dirty writes are present, and condition (\ref{c:3}) additionally avoids concurrent writes for transactions allocated to SI or SSI. In the proof we argue that if a condition is not satisfied, then at least one of the requirements of Definition~\ref{def:templ_split_sched} is not met.}

\smallskip
\noindent
{\bf Decision algorithm.}
The next proposition shows that it suffices to find a split schedule to decide template robustness.

\begin{propositionrep}
    \label{prop:robustness}
    Let $\templset$ be a set of transaction templates and let $\alloc^\templset$ be an allocation for $\templset$.
    The following are equivalent:
    \begin{itemize}
        \item $\templset$ is not robust against $\alloc^\templset$;
        \item there exists a template split schedule $\schedule$ for $\templset$ and $\alloc^\templset$ induced by a sequence of potentially conflicting quadruples $\seqpcq$ over $\templset$.
    \end{itemize}
\end{propositionrep}

\begin{proof}
    Due to Proposition~\ref{prop:split:non-conflict-serializable}, the existence of a template split schedule implies that $\templset$ is not robust against $\alloc^\templset$. The remainder of this proof is devoted to showing that if $\templset$ is not robust against $\alloc^\templset$, then there exists a template split schedule $\schedule$ for $\templset$ and $\alloc^\templset$ induced by a sequence of potentially conflicting quadruples $\seqpcq$ over $\templset$.
    {
    For this, we make use of Proposition~\ref{prop:conds:split} and show that, if $\templset$ is not robust against $\alloc^\templset$, a sequence of potentially conflicting quadruples $\seqpcq$ over $\templset$ can be constructed that makes the Conditions (\ref{c:1}) - (\ref{c:8}) from Proposition~\ref{prop:conds:split} true.

    First, it follows from Definition~\ref{def:template_robustness} that non-robustness against $\alloc^\templset$ means there exists a database $\db$, a set of transactions $\transset$ consistent with $\templset$ and $\db$, and an allocation $\alloc$ for $\transset$ consistent with $\alloc^\templset$, such that $\transset$ is not robust against $\alloc$. 
    Notice that, since every transaction $\trans[i] \in \transset$ is instantiated from a template $\templ[] \in \templset$ over $\db$, for every transaction $\trans[i]$ there is a mapping $\mu$ and template $\templ[]$ with $\mu(\templ[]) = \trans[i]$. In the remainder of the proof, we will consistently write $\mu_i(\templ[i])$ to mean $\trans[i]$ making its respective mapping and template explicit. 
    
    Now it follows from \cite[Theorem~3.2]{DBLP:conf/pods/VandevoortKN23} that a multiversion split schedule exists for $\transset$ based on a sequence of conflicting quadruples $c=(\mu_1(\templ[1]), \mu_1(o_1), \mu_2(p_2), \mu_2(\trans[2]), (\mu_2(\templ[2], o_2, p_3, \mu_3(\templ[3])), \ldots, (\mu_n(\templ[n]), \mu_n(o_m), \mu_1(p_1), \mu_1(\templ[1])$. 
    A conflicting quadruple (not to be confused with a \textit{potentially} conflicting quadruple) is a tuple $(\mu_i(\templ[i]), \mu_i(o_i), \mu_j(p_j), \mu_j(\templ[j]))$ with $(\templ[i], o_i, p_j, \templ[j])$ a potentially conflicting quadruple and with $\mu_i(o_i) = \mu_j(p_j)$. As a consequence, $(\templ[1], o_1, p_2, \templ[2]), (\templ[2], o_2, p_3, \templ[3]), \ldots, (\templ[n], o_n, p_1, \templ[1])$ is always a sequence of potentially conflicting quadruples $\seqpcq$. 
    A multiversion split schedule for $\transset$ and $\alloc$ based on $c$ is defined as a schedule of the form 
    \[\prefix{\mu_1(\templ[1])}{\mu_1(o_1)} \cdot \mu_2(\templ[2]) \cdot \ldots \cdot \mu_n(\templ[n]) \cdot \postfix{\mu_1(\templ[1])}{\mu_1(o_1)}\cdot \mu_{n+1}(\templ[n+1]) \cdot \ldots \cdot \mu_m(\templ[m]),\]
    with the following additional conditions:
    \begin{enumerate}[label=(\alph*)]
    \item \label{c:old:1} there is no operation in $T_1$ conflicting with an operation in any of the transactions $\trans[3], 
\ldots, \trans[n-1]$;
    \item \label{c:old:2} there is no write operation in $\prefix{\mu_1(\templ[1])}{o_1}$ ww-conflicting with a write operation in $\trans[2]$ or $\trans[n]$;
    \item \label{c:old:3} if $\alloc(\mu_1(\templ[1])) \in \{\si, \ssi\}$, then there is no write operation in $\postfix{\mu_1(o_1)}{\mu_1(\templ[1])}$ ww-conflicting with a write operation in $\mu_2(\templ[2])$ or $\mu_n(\templ[n])$;
    \item \label{c:old:4} $\mu_1(o_1)$ is rw-conflicting with $\mu_2(p_2)$;
    \item \label{c:old:5} $\mu_n(o_n)$ is rw-conflicting with $\mu_1(p_1)$ or $(\alloc(\mu_1(\templ[1])) = \rc \text{ and }\mu_1(o_1) <_{\templ[1]} \mu_1(p_1))$;
    \item \label{c:old:6} $\alloc(\mu_1(\templ[1])) \ne \ssi$ or $\alloc(\mu_2(\templ[2])) \ne \ssi$ or $\alloc(\mu_n(\templ[n])) \ne \ssi$; 
    \item \label{c:old:7} if $\alloc(\mu_1(\templ[1])) = \ssi$ and $\alloc(\mu_2(\templ[2]))=\ssi$, then there is no operation in $\mu_1(\templ[1])$ wr-conflicting with an operation in $\mu_2(\templ[2])$; and
    \item \label{c:old:8} if $\alloc(\mu_1(\templ[1])) = \ssi$ and $\alloc(\mu_n(\templ[n]))=\ssi$, then there is no operation in $\mu_1(\templ[1])$ rw-conflicting with an operation in $\mu_n(\templ[n])$.
    \end{enumerate}

    It remains to show that Conditions (\ref{c:1}) - (\ref{c:8}) from Proposition~\ref{prop:conds:split} are all true w.r.t.\@ the earlier mentioned sequence of potentially conflicting quadruples $\seqpcq$. The arguments follow rather immediately from Conditions \ref{c:old:1} - \ref{c:old:8}. 

    \smallskip\noindent
    For (\ref{c:1}), by contraposition: If there is an operation $o$ in $\templ[1]$ potentially conflicting with an operation $p$ of $\templ[j]$, with $j\in\{3, \ldots, n-1\}$ and with $\myvar{o}$ and $\myvar{p}$ connected in $\seqpcq$, then $\mu_1(o)$ must be conflicting with $\mu_j(p)$.

    \smallskip\noindent
    For (\ref{c:2}), by contraposition: If there is a write operation $o$ in $\prefix{\templ[1]}{o_1}$ potentially ww-conflicting with a write operation $p$ in $\templ[2]$ or in $\templ[n]$, where $\myvar{o}$ and $\myvar{p}$ are connected in $\seqpcq$, then $\mu_1(o)$ is a write operation in $\prefix{\mu_1(\templ[1])}{\mu_1(o_1)}$ ww-conflicting with $\mu_2(\templ[2])$ or $\mu_n(\templ[n])$. 

    \smallskip\noindent
    For (\ref{c:3}): If $\alloc^\templset(\templ[1]) \in \{\si, \ssi\}$ and there is a write operation $o$ in $\postfix{\templ[1]}{o_1}$ potentially ww-conflicting with a write operation $p$ in $\templ[2]$ or $\templ[n]$ with $\myvar{o}$ and $\myvar{p}$ connected in $\seqpcq$, then $\alloc(\mu_1(\templ[1])) \in \{\si, \ssi\}$ and write operation $\mu_1(o)$ in $\postfix{\mu_1(o_1)}{\mu_1(\templ[1])}$ is ww-conflicting with $\mu_2(\templ[2])$ or $\mu_n(\templ[n])$;

    \smallskip\noindent
    For (\ref{c:4}): Since $\mu_1(o_1)$ is rw-conflicting with $\mu_2(p_2)$, it must be the case that $o_{1}$ is potentially rw-conflicting with $p_{2}$.

    \smallskip\noindent
    For (\ref{c:5}): If $\mu_n(o_n)$ is rw-conflicting with $\mu_1(p_1)$ it follows directly that $o_{n}$ is potentially rw-conflicting with $p_{1}$. Otherwise, if $\alloc(\mu_1(\templ[1])) = \rc \text{ and }\mu_1(o_1) <_{\templ[1]} \mu_1(p_1)$ it follows directly that $\alloc^\templset(\templ[1]) = \rc$ and $o_1 <_{\templ[1]} p_1$;

    \smallskip\noindent
    For (\ref{c:6}): If $\alloc(\mu_1(\templ[1])) \ne \ssi$ then $\alloc^\templset(\templ[1]) \neq \ssi$. 
    If $\alloc(\mu_2(\templ[2])) \ne \ssi$ then $\alloc^\templset(\templ[2]) \neq \ssi$.
    Otherwise, it must be that case that $\alloc(\mu_n(\templ[n])) \ne \ssi$ and hence $\alloc^\templset(\templ[n]) \neq \ssi$.

    \smallskip\noindent
    For (\ref{c:7}):
    If $\alloc^\templset(\templ[1]) = \ssi$ and $\alloc^\templset(\templ[2]) = \ssi$, and 
    there is an operation $o$ in $\templ[1]$ potentially wr-conflicting with an operation $p$ in $\templ[2]$ with $\myvar{o}$ and $\myvar{p}$ connected in $\seqpcq$, then clearly $\alloc(\mu_1(\templ[1])) = \ssi$ and $\alloc(\mu_2(\templ[2]))=\ssi$ and $\mu_1(\templ[1])$ is wr-conflicting with $\mu_2(\templ[2])$, contradicting with Condition~\ref{c:old:7}.
    
    \smallskip\noindent
    For (\ref{c:8}): Assume towards a contradiction that $\alloc^\templset(\templ[1]) = \ssi$, $\alloc^\templset(\templ[n]) = \ssi$, and there is an operation $o$ in $\templ[1]$ potentially rw-conflicting with an operation $p$ in $\templ[n]$ with $\myvar{o}$ and $\myvar{p}$ connected in $\seqpcq$. Then, $\alloc(\mu_1(\templ[1])) = \ssi$, $\alloc(\mu_n(\templ[n]))=\ssi$, and $\mu_1(\templ[1])$ is rw-conflicting with $\mu_n(\templ[n])$, which contradicts with Condition~\ref{c:old:8}.
}
\end{proof}

\new{It readily follows from Proposition~\ref{prop:split:non-conflict-serializable} 
that a 
split schedule witnesses non-robustness. The reverse direction is more involved, and relies on the argument that we can extract a sequence of potentially conflicting quadruples satisfying the conditions in Proposition~\ref{prop:conds:split} from an arbitrary non-conflict-serializable schedule allowed under the allocation.}

Proposition~\ref{prop:conds:split} then 
offers a concrete way to find a split schedule via sequences of potentially conflicting quadruples.  However, a naive enumeration is not feasible, as these sequences can have an arbitrary length. Instead, we propose an algorithm that iterates over all possible choices for operations $o_1$ and $p_1$ in a template $\templ[1] \in \templset$, constructing a graph referred to as $\ptconflictgraph(o_1, p_1, \templ[1], h, \templset)$ with $h \in \{1,2\}$, which we will define next. Intuitively, the existence of a sequence $\seqpcq$ satisfying the conditions in Proposition~\ref{prop:conds:split} corresponds to reachability between specific nodes in this graph, as well as some additional conditions that the algorithm will verify separately.

Let $\templset$ be a set of transaction templates, $o_1$ and $p_1$ two (not necessarily different) operations occurring in a template $\templ[1] \in \templset$, and let $h \in \{1,2\}$. The directed graph $\ptconflictgraph(o_1, p_1, \templ[1], h, \templset)$ has nodes of the form $(\templ[], o, c, k)$ for all $\templ[] \in \templset$, $o\in \templ[]$, $c \in \{O,P,N\}$ and $k \in \{\text{in}, \text{out}\}$ satisfying the following conditions:
\begin{enumerate}
    \item if $c = O$, there is no operation $o_1' \in \templ[1]$ over the same variable as $o_1$ such that $o_1'$ is potentially conflicting with an operation $o' \in \templ[]$ and $\myvar{o'} = \myvar{o}$;
    \item if $c = P$, there is no operation $o_1' \in \templ[1]$ over the same variable as $p_1$ such that $o_1'$ is potentially conflicting with an operation $o' \in \templ[]$ and $\myvar{o'} = \myvar{o}$.
\end{enumerate}
Intuitively, the value $h$ indicates whether $\myvar{o_1}$ and $\myvar{p_1}$ are connected in the sequence $\seqpcq$ that will be constructed by the algorithm, where $h = 1$ indicates that they are connected and $h = 2$ indicates that they are not. For each node $(\templ[], o, c, k)$, the value of $c$ indicates that $o$ is connected to $o_1$ ($c = O$), to $p_1$ ($c = P$), or to neither ($c = N$) in $\seqpcq$. Lastly, the value of $k$ indicates whether $o$ is an incoming or outgoing operation in a quadruple in $\seqpcq$.
The previous two conditions over these nodes then guarantee that the sequence $\seqpcq$ constructed by the algorithm satisfies Condition~\ref{c:1} in Proposition~\ref{prop:conds:split}.

The graph $\ptconflictgraph(o_1, p_1, \templ[1], h, \templset)$ contains an edge from a node $(\templ[], o, c, k)$ to a node $(\templ[]', o', c', k')$ if and only if one of the following conditions hold:
\begin{enumerate}
    \setcounter{enumi}{2}
    \item \label{c:g:3} $k = \text{out}$,  $k' = \text{in}$, $c = c'$, and $o$ is potentially conflicting with $o'$;
    \item \label{c:g:4} $k = \text{in}$,  $k' = \text{out}$, $\templ[] = \templ[]'$, $\myvar{o} \neq \myvar{o'}$ and $(c, c') \in \{(O,P), (O,N), (N,N), (N,P)\}$;
    \item \label{c:g:5} $k = \text{in}$,  $k' = \text{out}$, $\templ[] = \templ[]'$, $\myvar{o} = \myvar{o'}$ and $c = c'$; or
    \item \label{c:g:6} $k = \text{in}$,  $k' = \text{out}$, $\templ[] = \templ[]'$, $\myvar{o} = \myvar{o'}$, $c = O$, $c' = P$ and $h = 1$.
\end{enumerate}
By construction, these edges propagate connectedness with respect to $\myvar{o_1}$ and $\myvar{p_1}$ in the sequence $\seqpcq$ that will be constructed by the algorithm. Condition~(\ref{c:g:3}) requires potentially conflicting operations for each quadruple in $\seqpcq$, while maintaining connectedness (recall that the variables over which the operations in a quadruple are defined are always connected). Within a template, connectedness with $o_1$ or $p_1$ only propagates if the variables are the same (Condition~(\ref{c:g:4}) and Condition~(\ref{c:g:5})). Condition~(\ref{c:g:6}) covers the special case when we assume that $\myvar{o_1}$ and $\myvar{p_1}$ are connected (i.e., $h = 1$). In that case, every variable connected to $\myvar{o_1}$ is connected to $\myvar{p_1}$ as well.

The template robustness algorithm is displayed as Algorithm~\ref{alg:templ_rob}.
The algorithm iterates over all possible choices for operations in $\templ[1]$, $\templ[2]$ and $\templ[n]$, tracking connectedness of $\myvar{o_2}$ and $\myvar{p_n}$ with respect to $\myvar{o_1}$ and $\myvar{p_1}$. For each such choice, the algorithm then verifies whether $p_n$ is reachable from $o_2$ in the $\ptconflictgraph$, thereby witnessing the existence of a sequence of potentially conflicting quadruples satisfying Condition~(\ref{c:1}) of Proposition~\ref{prop:conds:split}, followed by the verification of the remaining conditions. 
\inConfVersion{The implementation of \textit{Reachable} and \textit{ValidSchedule} are straightforward and provided in detail in \cite{fullversion_repmila}.}
\inFullVersion{The details of \textit{Reachable} and \textit{ValidSchedule} are provided in Algorithms~\ref{alg:func:reachable} and~\ref{alg:func:validschedule}, respectively.}

\begin{algorithm}[t]
    \caption{Deciding template robustness}
    \label{alg:templ_rob}
    \SetKwInOut{Input}{Input}
    \SetKwInOut{Output}{Output}
    \SetKwComment{Comment}{//}{}
    \SetKwProg{Fn}{Function}{}{}
    \Input{Set of transaction templates $\templset$ and template allocation $\alloc^\templset$ for $\templset$}
    \Output{\emph{True} iff $\templset$ is robust against $\alloc^\templset$}
    \BlankLine
    
    \ForEach{$o_1, p_1 \in \templ[1]$ with $\templ[1] \in \templset$}
    {
        \lIf*{$\myvar{o_1} = \myvar{p_1}$}{$H := \{1\}$} \lElse{$H := \{1,2\}$}
        \ForEach{$h \in H$}
        {
            $G := \ptconflictgraph(o_1, p_1, \templ[1], h, \templset)$\;
            $TC := \text{transitive closure of } G$\;
            \ForEach{$o_2, p_2 \in \templ[2]$; $o_n, p_n \in \templ[n]$ with $\templ[2], \templ[n] \in \templset$}
            {
                \If{$o_1$ not potentially conflicting with $p_2$ \textbf{or}\\
                \phantom{\textbf{If} }$o_n$ not potentially conflicting with $p_1$}
                {
                    \textbf{continue}\;
                }
                \lIf*{$\myvar{o_2} = \myvar{p_2}$}{$C_{o2} := \{O\}$} \lElse{$C_{o2} := \{N,P\}$}
                \lIf*{$\myvar{o_n} = \myvar{p_n}$}{$C_{pn} := \{P\}$} \lElse{$C_{pn} := \{N,O\}$}
                \ForEach{$c_{o2} \in C_{o2}$, $c_{pn} \in C_{pn}$}
                {
                    \If{Reachable($\templ[2], o_2, p_2, c_{o2},$\\
                    \mbox{}\phantom{\textbf{if} \textit{Reachable(}}$\templ[n], o_n, p_n, c_{pn}, h, TC$) \textbf{and}\\
                    \mbox{}\phantom{\textbf{if} }ValidSchedule($\tau_1, o_1, p_1, \tau_2, o_2, p_2, c_{o2},$\\
                    \mbox{}\phantom{\textbf{if} ValidSchedule(}$\tau_n, o_n, p_n, c_{pn}, h, \alloc^\templset$)}
                    {
                        \Return{\emph{False}}\;
                    }
                }
            }
        }
    }
    \Return{\emph{True}}\;
\end{algorithm}

\inFullVersion{
\begin{algorithm}[t]
    \SetKwInOut{Input}{Input}
    \SetKwInOut{Output}{Output}
    \SetAlgoLined


    \tcc{Case $n=2$:}
    \If{$\templ[2] = \templ[n]$ \textbf{and} $o_2 = o_n$ \textbf{and} $p_2 = p_n$}
    {
        \tcc{Valid if $o_2$ is connected to $p_1$, and $p_2 = p_n$ is connected to $o_1$:}
        \If{$c_{o2} = P$ \textbf{and} $c_{pn} = O$}
        {
            \Return{True}\;
        }
        \tcc{Special case: $o_2$ and $p_2$ are over same variable, implying that $o_1$ and $p_1$ are connected ($h = 1$):}
        \If{$h = 1$ \textbf{and} $c_{o2} = O$ \textbf{and} $c_{pn} = P$}
        {
            \Return{True}\;
        }
    }

    \tcc{Case $n=3$:}
    \If{$o_2$ potentially conflicting with $p_n$}
    {
        \tcc{Valid if they agree on connectedness:}
        \If{$c_{o2} = c_{pn}$}
        {
            \Return{True}\;
        }
        \tcc{Special case: $o_2$ connected to $o_1$, and $p_n$ connected to $p_1$, but $o_1$ and $p_1$ are connected ($h = 1$):}
        \If{$h = 1$ \textbf{and} $c_{o2} = O$ \textbf{and} $c_{pn} = P$}
        {
            \Return{True}\;
        }
    }
    
    \tcc{Case $n>3$:}
    \ForEach{edge $(n, n')$ in $TC$}
    {
        \textbf{Let} $n = (\templ[], o, c, k)$ \textbf{and} $n' = (\templ[]', o', c', k')$\;
        \If{$k = \text{in}$ \textbf{and} $k'=\text{out}$ \textbf{and} $c = c_{o2}$ \textbf{and} $c' = c_{pn}$ \textbf{and} \\
        \phantom{\textbf{If} }$o$ is potentially conflicting with $o_2$ \textbf{and}\\
        \phantom{\textbf{If} }$o'$ is potentially conflicting with $p_n$}
        {
            \Return{True}\;
        }
    }

    \Return{False}\;

    \caption{\label{alg:func:reachable} Reachable($\templ[2], o_2, p_2, c_{o2}, \templ[n], o_n, p_n, c_{pn}, h, TC$)}
\end{algorithm}
}

\inFullVersion{
\begin{algorithm}[t]
    \SetKwInOut{Input}{Input}
    \SetKwInOut{Output}{Output}
    \SetKwProg{Fn}{Function}{}{}
    \SetAlgoLined


    \tcc{Condition (\ref{c:2}):}
    \ForEach{$o_1' \in \prefix{\templ[1]}{o_1}$}
    {
        \ForEach{$p_2' \in \templ[2]$}
        {
            \If{$o_1'$ potentially ww-conflicting with $p_2'$ \textbf{and} $\myvar{o_1'}$ connected to $\myvar{p_2'}$}
            {
                \Return{False}\;
            }
        }
        \ForEach{$p_n' \in \templ[n]$}
        {
            \If{$o_1'$ potentially ww-conflicting with $p_n'$ \textbf{and} $\myvar{o_1'}$ connected to $\myvar{p_n'}$}
            {
                \Return{False}\;
            }
        }
    }

    \BlankLine
    \tcc{Condition (\ref{c:3}):}
    \If{$\alloc^\templset(\templ[1]) \in \{\si, \ssi\}$}
    {
        \ForEach{$o_1' \in \postfix{\templ[1]}{o_1}$}
        {
            \ForEach{$p_2' \in \templ[2]$}
            {
                \If{$o_1'$ potentially ww-conflicting with $p_2'$ \textbf{and} $\myvar{o_1'}$ connected to $\myvar{p_2'}$}
                {
                    \Return{False}\;
                }
            }
            \ForEach{$p_n' \in \templ[n]$}
            {
                \If{$o_1'$ potentially ww-conflicting with $p_n'$ \textbf{and} $\myvar{o_1'}$ connected to $\myvar{p_n'}$}
                {
                    \Return{False}\;
                }
            }
        }
    }

    \BlankLine
    \tcc{Condition (\ref{c:4}):}
    \If{$o_1$ not potentially rw-conflicting with $p_2$}
    {
        \Return{False}\;
    }

    \BlankLine
    \tcc{Condition (\ref{c:5}):}
    \If{$o_n$ not potentially rw-conflicting with $p_1$}
    {
        \If{$\alloc^\templset(\templ[1]) \in \{\si, \ssi\}$ \textbf{or} $p_1 <_{\templ[1]} o_1$}
        {
            \Return{False}\;
        }
    }

    \BlankLine
    \tcc{Condition (\ref{c:6}):}
    \If{$\alloc^\templset(\templ[1]) = \alloc^\templset(\templ[2]) = \alloc^\templset(\templ[n]) = \ssi$}
    {
        \Return{False}\;
    }

    \BlankLine
    \tcc{Condition (\ref{c:7}):}
    \If{$\alloc^\templset(\templ[1]) = \ssi$ \textbf{and}$\alloc^\templset(\templ[2]) = \ssi$}
    {
        \ForEach{$o_1' \in \templ[1]$}
        {
            \ForEach{$p_2' \in \templ[2]$}
            {
                \If{$o_1'$ potentially wr-conflicting with $p_2'$ \textbf{and} $\myvar{o_1'}$ connected to $\myvar{p_2'}$}
                {
                    \Return{False}\;
                }
            }
        }
    }

    \BlankLine
    \tcc{Condition (\ref{c:8}):}
    \If{$\alloc^\templset(\templ[1]) = \ssi$ \textbf{and}$\alloc^\templset(\templ[n]) = \ssi$}
    {
        \ForEach{$o_1' \in \templ[1]$}
        {
            \ForEach{$p_n' \in \templ[n]$}
            {
                \If{$o_1'$ potentially rw-conflicting with $p_n'$ \textbf{and} $\myvar{o_1'}$ connected to $\myvar{p_n'}$}
                {
                    \Return{False}\;
                }
            }
        }
    }
        
    \Return{True}\;

    \caption{\label{alg:func:validschedule} ValidSchedule($\tau_1, o_1, p_1, \tau_2, o_2, p_2, c_2,$\\
    \phantom{\textbf{Algorithm 3:} Validschedule(}$\tau_n, o_n, p_n, c_n, h, \alloc^\templset$)}
\end{algorithm}
}

\begin{theoremrep}\label{thm:algo:rob:ptime}
    Let $\templset$ be a set of templates and $\alloc^\templset$ a template allocation for $\templset$.
    Algorithm~\ref{alg:templ_rob} decides whether\/ 
    $\templset$ is robust against $\alloc^\templset$ in time polynomial in the size of\/ $\templset$.
\end{theoremrep}

\begin{proof}
The correctness of the algorithm follows from Proposition~\ref{prop:conds:split} and Proposition~\ref{prop:robustness}. In particular, the algorithm verifies robustness by testing whether a template split schedule exists, where existence of such a schedule is based on the conditions provided in Proposition~\ref{prop:conds:split}. The algorithm iterates over all possible choices for operations $o_1$ and $p_1$ in $\templ[1]$, constructing the graph $\ptconflictgraph(o_1, p_1, \templ[1], h, \templset)$ for each choice. For each such graph, the algorithm then verifies whether $p_n$ is reachable from $o_2$ in the transitive closure of the graph, thereby witnessing the existence of a sequence of potentially conflicting quadruples satisfying Condition~(\ref{c:1}) of Proposition~\ref{prop:conds:split}, followed by the verification of the remaining conditions.

Towards the complexity analysis, note in particular that the constructed $\ptconflictgraph$ is polynomial in the size of the input. Indeed, the number of nodes in the graph is linear in the number of operations in $\templset$, and computing the transitive closure over this graph is therefore possible in polynomial time. It is now straightforward to see that the algorithm as a whole, including functions \textit{Reachable} and \textit{ValidSchedule}, runs in time polynomial in the size of the input.
\end{proof}

\subsection{Finding Lowest Robust Allocations}
\label{sec:allocation}

We first show that there always exists a unique lowest robust allocation and then present a polynomial time algorithm to find it.

As discussed in Section~\ref{sec:SBexample}, we assume a total order $\rc < \si < \ssi$ over the three considered isolation levels, expressing our preference for lower isolation levels over higher ones. In the following, let $\alloc^\templset_1$ and $\alloc^\templset_2$ be two template allocations for a set of templates $\templset$. We write $\alloc^\templset_1 \leq \alloc^\templset_2$ if $\alloc^\templset_1(\templ[]) \leq \alloc^\templset_2(\templ[])$ for all $\templ[] \in \templset$. Furthermore, we write $\alloc^\templset_1 < \alloc^\templset_2$ if $\alloc^\templset_1 \leq \alloc^\templset_2$ and there exists a template $\templ[] \in \templset$ such that $\alloc^\templset_1(\templ[]) < \alloc^\templset_2(\templ[])$.
For an isolation level $\isolationlevel$, we denote by $\alloc^\templset_1[\templ[] \mapsto \isolationlevel]$ the template allocation for $\templset$ where $\alloc^\templset_1[\templ[] \mapsto \isolationlevel](\templ[]) = \isolationlevel$ and $\alloc^\templset_1[\templ[] \mapsto \isolationlevel](\templ[]') = \alloc^\templset_1(\templ[]')$ for all other templates $\templ[]' \in \templset$. That is, the template allocation derived from $\alloc^\templset_1$ by setting the isolation level of $\templ[]$ to $\isolationlevel$, while leaving all other templates unchanged.

\begin{definition}
    Let $\templset$ be a set of templates robust against a template allocation $\alloc^\templset_1$ for $\templset$. Then, $\alloc^\templset_1$ is \emph{lowest} if there is no other allocation $\alloc^\templset_2$ for $\templset$ such that $\alloc^\templset_2 < \alloc^\templset_1$ and $\templset$ is robust against $\alloc^\templset_2$.
\end{definition}

The next propositions extend some results for allocations for transactions presented in~\cite{DBLP:conf/pods/VandevoortKN23} towards template allocations:

\begin{propositionrep}\label{prop:robustallocations}
    Let $\templset$ be a set of templates, and let $\alloc^\templset_1$ and $\alloc^\templset_2$ be two template allocations for $\templset$.
    \begin{enumerate}
        \item If $\alloc^\templset_1 \leq \alloc^\templset_2$ and $\templset$ is robust against $\alloc^\templset_1$, then $\templset$ is robust against $\alloc^\templset_2$.
        \item \label{prop:robustallocations:item2} If $\templset$ is robust against $\alloc^\templset_1$ and $\alloc^\templset_2$, then $\templset$ is robust against $\alloc^\templset_2[\templ[] \mapsto \alloc^\templset_1(\templ[])]$ for every $\templ[] \in \templset$.
    \end{enumerate}
\end{propositionrep}

\begin{proof}
{
    \emph{(1)} The proof is by contraposition. Assume $\alloc^\templset_1 \leq \alloc^\templset_2$ and $\templset$ is not robust against $\alloc^\templset_2$. We argue that $\templset$ is not robust against $\alloc^\templset_1$.
    According to Proposition~\ref{prop:conds:split} and Proposition~\ref{prop:robustness}, there is a template split schedule $\schedule$ for $\templset$ and $\alloc^\templset_2$ induced by a sequence of potentially conflicting quadruples over $\templset$ with $\seqpcq$ having conditions~(\ref{c:1})-(\ref{c:8}) of Proposition~\ref{prop:conds:split}.
    To conclude that $\templset$ is not robust against $\alloc^\templset_1$, we need to show that conditions~(\ref{c:1})-(\ref{c:8}) of Proposition~\ref{prop:conds:split} also apply to $\templset$, $\alloc^\templset_1$ and $\seqpcq$. 

    To see that these conditions indeed apply to $\alloc^\templset_1$, we make the following observation: Since $\alloc_1^\templset \leq \alloc^\templset_2$, we have
    \begin{itemize}
        \item $\alloc^\templset_2(\templ[])=\mvrc \Rightarrow \alloc^\templset_1(\templ[])=\mvrc$;
        \item $\alloc^\templset_1(\templ[]) \in \{\si, \ssi\} \Rightarrow \alloc^\templset_2(\templ[]) \in \{\si, \ssi\}$; and
        \item $\alloc^\templset_1(\templ[])=\ssi \Rightarrow \alloc^\templset_2(\templ[])=\ssi$
    \end{itemize}
    for all $\templ[] \in \templset$. Now, it becomes straightforward to verify that if the conditions in Proposition~\ref{prop:conds:split} hold for $\alloc^\templset_2$, then they must hold for $\alloc^\templset_1$ as well. 
    With this result in place, it follows from Proposition~\ref{prop:conds:split} that a template split schedule for $\templ$ and $\alloc^\templset_1$ induced by $\seqpcq$ exists. The desires result, that $\templset$ is not robust against $\alloc^\templset_1$, follows from Proposition~\ref{prop:robustness}.

    \medskip
    \noindent
    \emph{(2)}
    The proof is by contraposition. Assume $\templset$ is not robust against $\alloc^\templset_3 = \alloc^\templset_2[\templ[] \mapsto \alloc^\templset_1(\templ[])]$ for some $\templ[] \in \templset$. We argue by case on $\templ[]$ that $\templset$ is not robust against $\alloc^\templset_1$ or $\alloc^\templset_2$.

    According to Proposition~\ref{prop:conds:split} and Proposition~\ref{prop:robustness}
    Conditions~(\ref{c:1}) - (\ref{c:8}) of Proposition~\ref{prop:conds:split} are true for $\templset$ and $\alloc^\templset_3$ 
    induced by some sequence of potentially conflicting quadruples $\seqpcq = (\templ[1], o_1, p_2, \templ[2]), (\templ[2], o_2, p_3, \templ[3]), \ldots, (\templ[n], o_n, p_1, \templ[1])$.
    If $\templ[] \not \in \{\templ[1], \templ[2], \templ[n]\}$, all conditions of Proposition~\ref{prop:conds:split} apply to $\templset$, $\alloc^\templset_2$ and $\seqpcq$ as well, since the conditions of Proposition~\ref{prop:conds:split} only consider the allocated isolation level for templates $\templ[1]$, $\templ[2]$ and $\templ[n]$, and since $\alloc^\templset_3(\templ[]) = \alloc^\templset_2(\templ[])$ for $\templ[] \in \{\templ[1], \templ[2], \templ[n]\}$. 
    Hence, in this case it is immediate that $\templset$ is not robust against $\alloc^\templset_2$.

    Otherwise, if $\templ[] \in \{\templ[1], \templ[2], \templ[n]\}$, assume towards a contradiction that $\templset$ is robust against both $\alloc^\templset_1$ and $\alloc^\templset_2$. Let $\allocb^\templset_1$, $\allocb^\templset_2$ and $\allocb^\templset_3$ be the allocations for $\templset$ obtained by taking $\allocb^\templset_1(\templ[i]) = \alloc^\templset_1(\templ[i])$, $\allocb^\templset_2(\templ[i]) = \alloc^\templset_2(\templ[i])$ and $\allocb^\templset_3(\templ[i]) = \alloc^\templset_2(\templ[i])$
    if $\templ[i] \in \{\templ[1], \templ[2], \templ[n]\}$ and $\allocb^\templset_1(\templ[i]) = \allocb^\templset_2(\templ[i]) = \allocb^\templset_3(\templ[i]) = \ssi$ otherwise. By construction, $\alloc^\templset_1 \leq \allocb^\templset_1$, $\alloc^\templset_2 \leq \allocb^\templset_2$ and $\alloc^\templset_3 \leq \allocb^\templset_3$. By (1) we conclude that 
    $\templset$ is robust against $\allocb^\templset_1$ and $\allocb^\templset_2$ as well.
    Notice furthermore that $\allocb^\templset_3 = \allocb^\templset_2[\templ[] \mapsto \allocb^\templset_1(\templ[])]$ and hence that the conditions of Proposition~\ref{prop:conds:split} (which are assumed true for 
    $\templset$, $\alloc^\templset_3$ and $\seqpcq$) also apply to $\templset$, $\allocb^\templset_3$ and $\seqpcq$, since $\allocb^\templset_3$ agrees with $\alloc^\templset_3$ on $\templ[1]$, $\templ[2]$ and $\templ[n]$. 
    In other words, $\templset$ is not robust against $\allocb^\templset_3$.
    
    Since $\templset$ is robust against $\allocb^\templset_1$ and $\allocb^\templset_2$ but not against $\allocb^\templset_3$, it follows from (1) that 
    $\allocb^\templset_1 \not \leq \allocb^\templset_3$ and $\allocb^\templset_2 \not \leq\allocb^\templset_3$.
    By construction of $\allocb^\templset_1$, $\allocb^\templset_2$ and $\allocb^\templset_3$, in particular the fact that $\allocb^\templset_2$ and $\allocb^\templset_3$ only differ on $\templ[]$, we therefore conclude that 
    $\allocb^\templset_1(\templ[]) < \allocb^\templset_2(\templ[])$
     (as otherwise $\allocb^\templset_2 \leq \allocb^\templset_3$).
    Furthermore, 
    there exists a template $\templ[]' \in \{\templ[1], \templ[2], \templ[n]\}$ with $\templ[]' \neq \templ[]$ and $\allocb^\templset_2(\templ[]') < \allocb^\templset_1(\templ[]')$.
    Indeed, if for each such $\templ[]'$ we had $\allocb^\templset_1(\templ[]') \leq \allocb^\templset_2(\templ[]')$ and hence $\allocb^\templset_1(\templ[]') \leq \allocb^\templset_3(\templ[]')$ (since $\allocb^\templset_2(\templ[]') = \allocb^\templset_3(\templ[]')$), then $\allocb^\templset_1(\templ[]) = \allocb^\templset_3(\templ[])$ and $\allocb^\templset_1(\templ[]'') = \allocb^\templset_3(\templ[]'')$ for each $\templ[]'' \not\in \{\templ[1], \templ[2], \templ[n]\}$, implying $\allocb^\templset_1 \leq \allocb^\templset_3$, which would be a contradiction against the earlier proven facts that $\templset$ is robust against $\allocb^\templset_1$ but not $\allocb^\templset_3$.

    Since $\allocb^\templset_1(\templ[]) < \allocb^\templset_2(\templ[])$, we have 
    $\allocb^\templset_1(\templ[]) \neq \ssi$, and analogously $\allocb^\templset_2(\templ[]') < \allocb^\templset_1(\templ[]')$ implies 
    $\allocb^\templset_2({\templ[]'}) \neq \ssi$.
    In the remainder of the proof, we argue that, for each choice of $\templ[] \in \{\templ[1], \templ[2], \templ[m]\}$, $\templset$ is not robust against either $\allocb^\templset_1$ or $\allocb^\templset_2$, thereby leading to the desired contradiction.

    If $\templ[] = \templ[1]$, we have $\allocb^\templset_3(\templ[1]) = \allocb^\templset_1(\templ[1])$ and $\allocb^\templset_1(\templ[1]) \neq \ssi$. 
    As a result, the conditions of Proposition~\ref{prop:conds:split} apply to $\templset$, $\allocb^\templset_1$ and $\seqpcq$. 
    By Proposition~\ref{prop:robustness}, $\templset$ is therefore not robust against $\allocb^\templset_1$.

    If $\templ[] = \templ[2]$, we have $\allocb^\templset_3(\templ[1]) = \allocb^\templset_2(\templ[1])$ and $\templ[]' \in \{\templ[1], \templ[n]\}$, thereby implying that $\allocb^\templset_2(\templ[1]) \neq \ssi$ or $\allocb^\templset_2(\templ[n]) \neq \ssi$.
    We argue that the conditions of Proposition~\ref{prop:conds:split} are true for $\templset$, $\allocb^\templset_2$ and $\seqpcq$, thereby implying that $\templset$ is not robust against $\allocb^\templset_2$. 
    Notice that Condition~\ref{c:7} of Proposition~\ref{prop:conds:split} is the only condition that is not immediate by the observations above. Towards a contradiction, assume it does not hold. That is, $\allocb^\templset_2(\templ[1]) = \ssi$, $\allocb^\templset_2(\templ[2]) = \ssi$ and there is an operation $o$ in $\templ[1]$ potentially wr-conflicting with an operation $p$ in $\templ[2]$ with $\myvar{o}$ and $\myvar{p}$ connected in $\seqpcq$. 
    The latter means that in every template split schedule $\schedule$ for $\templset$ and $\alloc^\templset$ induced by $\seqpcq$ (see defined Definition~\ref{def:template}), there is an rw-antidependency from $\mu_2(\tau_2)$ to $\mu_1(\tau_1)$.
    For this, particularly notice that $\schvf(\mu_2(p)) \schvord \mu_1(o)$ because the commit of $\mu_1(\tau_1)$ occurs after $\mu_2(p)$ in $\schedule$ and $\alloc(\mu(\tau_2)) = \ssi$.  
    At the same time, and following the same reasoning, it follows from condition (\ref{c:4}) that $o_1$ is potentially rw-conflicting with $p_2$, hence we also always have an rw-antidependency from $\mu_1(\tau_1)$ to $\mu_2(\tau_2)$.
    Since both transactions $\mu_1(\tau_1)$ and $\mu_2(\tau_2)$ are concurrent, we conclude that every template split schedule for $\templset$ and $\alloc^\templset$ induced by $\seqpcq$
    has a dangerous structure in $\schedule$ and hence is not allowed by $\alloc$. This observation contradicts with our assumption that $\seqpcq$ has the conditions of Proposition~\ref{prop:conds:split} implying that a template split schedule for $\templset$ and $\alloc^\templset$ induced by $\seqpcq$ exists.

    By construction of $\seqpcq$, this pair of potentially conflicting operations implies a potential rw-antidependency from $\templ[2]$ to $\templ[1]$. 
    According to $(\dagger)$, we have $\templ[2] = \templ[n]$. But then we have $\allocb^\templset_2(\templ[2]) = \ssi$ and $\allocb^\templset_2(\templ[n]) \neq \ssi$, leading to the desired contradiction.
    
    The case where $\templ[] = \templ[n]$ is analogous to the previous case. We have $\allocb^\templset_3(\templ[1]) = \allocb^\templset_2(\templ[1])$ and $\templ[]' \in \{\templ[1], \templ[2]\}$, thereby implying that $\allocb^\templset_2(\templ[1]) \neq \ssi$ or $\allocb^\templset_2(\templ[2]) \neq \ssi$.
    Again, we argue that the conditions of Proposition~\ref{prop:conds:split} are true for $\templset$, $\allocb^\templset_2$ and $\seqpcq$, thereby implying that $\templset$ is not robust against $\allocb^\templset_2$.
    In this case, Condition~\ref{c:8} of Proposition~\ref{prop:conds:split} is the only one that is not immediate by the observations above. Towards a contradiction, assume it does not hold. That is, $\allocb^\templset_2(\templ[1]) = \ssi$, $\allocb^\templset_2(\templ[n]) = \ssi$ and there is an operation in $\templ[1]$ potentially rw-conflicting with an operation in $\templ[n]$. Following analogous reasoning as for the previous case, we can conclude that every template split schedule for $\templset$ and $\alloc^\templset$ induced $\seqpcq$ contains a dangerous structure and hence is not allowed by $\alloc$, which contradicts, as before, with Proposition~\ref{prop:conds:split} and our assumptions about $\seqpcq$.
    } 
\end{proof}

\begin{propositionrep}
    There exists a unique lowest template allocation for every set of templates $\templset$.
\end{propositionrep}

\begin{proof}
    {
        Suppose towards a contradiction that there are two different lowest template allocations $\alloc^\templset_1$ and $\alloc^\templset_2$.
        As $\alloc^\templset_1$ and $\alloc^\templset_2$ are different, there exists a template $\templ[]\in\templset$ such that $\alloc^\templset_1(\templ[])\neq\alloc^\templset_2(\templ[])$. 
        Without loss of generality, we assume $\alloc^\templset_1(\templ[])<\alloc^\templset_2(\templ[])$.
        By Proposition~\ref{prop:robustallocations}~(\ref{prop:robustallocations:item2}), $\templset$ is robust against $\alloc^\templset_2[\templ[]\mapsto \alloc^\templset_1(\templ[])]$. But then $\alloc^\templset_2[\templ[]\mapsto \alloc^\templset_1(\templ[])] < \alloc^\templset_2$, which contradicts with the assumption that $\alloc^\templset_2$ is lowest.}
\end{proof}

Algorithm~\ref{alg:optimalallocation} outlines the procedure for determining the unique lowest robust allocation. Initially, all templates are assigned the allocation $\ssi$. The algorithm then iterates through each template, evaluating whether the associated isolation level can be safely reduced using Algorithm~\ref{alg:templ_rob} for the robustness test.

\begin{algorithm}[t]
    \SetKwInOut{Input}{Input}
    \SetKwInOut{Output}{Output}
    \SetAlgoLined

    \Input{\ Set of templates $\templset$}
    \Output{\ Lowest robust template allocation $\alloc^\templset$ for $\templset$}
    \BlankLine
    \For{$\templ[] \in \templset$}{
        $\alloc^\templset[\templ[]] := \ssi$\;
    }
    \BlankLine
    \For{$\templ[] \in \templset$}{
        \If{$\templset$ is robust against $\alloc^\templset[\templ[] \mapsto \mvrc]$}{
            $\alloc^\templset := \alloc^\templset[\templ[] \mapsto \mvrc]$\;
        }
        \ElseIf{$\templset$ is robust against $\alloc^\templset[\templ[] \mapsto \si]$}{
            $\alloc^\templset := \alloc^\templset[\templ[] \mapsto \si]$\;
        }
    }
    \Return $\alloc^\templset$\;

    \caption{\label{alg:optimalallocation} Computing the lowest robust allocation.}
\end{algorithm}

\begin{theoremrep}
    For a set of templates $\templset$,
    Algorithm~\ref{alg:optimalallocation} computes the
    unique lowest robust template allocation in time polynomial in the size of\/ $\templset$.
\end{theoremrep}

\begin{proof}
    By assumption $\templset$ is robust against the allocation $\alloc^\templset_{\ssi}$ that maps all transactions to \ssi.
    Algorithm~\ref{alg:optimalallocation} then refines this allocation by assigning the minimal isolation level to each template leading to a lowest robust allocation.
    The correctness follows by repeated application of Proposition~\ref{prop:robustallocations}(\ref{prop:robustallocations:item2}). It follows in particular from Proposition~\ref{prop:robustallocations}(\ref{prop:robustallocations:item2}) that for every robust allocation $\alloc^\templset$ for $\templset$ (including $\alloc^\templset_{\ssi}$) there is a sequence
    of allocations $\alloc^\templset_1 < \alloc^\templset_2 < \ldots <\alloc^\templset_k < \alloc^\templset$ with $\alloc^\templset_1$ denoting the unique lowest allocation for $\transset$ and $\alloc^\templset_i = \alloc^\templset_{i+1}[\templ[]\to \alloc^\templset_{i}(\templ[])]$ for every $i \in [1,k]$. The polynomial time complexity follows directly from Theorem~\ref{thm:algo:rob:ptime}.
\end{proof}


\section{Related Work}
\label{sec:relwork}


Shortly after the initial papers that defined serializability and studied its theoretical complexity \cite{DBLP:journals/jacm/Papadimitriou79b}, the IBM System R team published an account of weaker isolation levels and locking-based algorithms that achieved those by releasing shared locks early, or even not taking them at all \cite{DBLP:conf/ds/GrayLPT76}. Much later, the multiversion Snapshot Isolation mechanism was described, and shown to allow some non-serializable executions, despite avoiding all the anomalies mentioned in the SQL specification \cite{DBLP:conf/sigmod/BerensonBGMOO95}. The serializable multiversion SSI mechanism was proposed by Cahill \cite{DBLP:journals/tods/CahillRF09} and was implemented (with optimizations) in PostgreSQL \cite{PortsGrittner2012}.

There has been a variety of approaches used to define isolation properties abstractly. For definitions and proofs that mechanisms achieve serializability, theory was developed especially by Bernstein and Goodman, including for multiversion and even distributed protocols \cite{DBLP:journals/tods/BernsteinG83, DBLP:books/aw/BernsteinHG87}. For defining lower isolation, Gray et al. \cite{DBLP:conf/ds/GrayLPT76} and the later SQL specification used the notion of anomalies, that is patterns of read and write operations that occur in the system's histories and can lead to situations that do not happen in serial executions. Berenson et al. \cite{DBLP:conf/sigmod/BerensonBGMOO95} showed that this approach was inadequate to deal with Snapshot Isolation, and proposed some alternatives that were later seen as also inadequate. 
Adya developed a theory framework for defining these isolation levels abstractly, based on graphs showing dependency edges between operations\cite{adya99, DBLP:conf/icde/AdyaLO00}. Much ongoing work has built on Adya's style. Cerone et al.  \cite{DBLP:conf/concur/Cerone0G15} and Crooks et al. \cite{DBLP:conf/podc/CrooksPAC17} offer different approaches based on abstract state or a ``client-centric'' definition style.

Shasha et al. introduced the approach of showing conditions which can prove robustness, showing when serializable execution is ensured despite using mechanisms that in general allow non-serializable behavior (in this case, dividing a transaction into segments separated by COMMIT operations)  \cite{DBLP:journals/tods/ShashaLSV95}. Fekete et al. gave a theory that could prove robustness for Snapshot Isolation \cite{DBLP:journals/tods/FeketeLOOS05}. This paper also introduced the technique of making an application robust through modifying application code by ``promoting'' a read to also do an identity update of the item.
Alomari et al. examined performance comparisons of promotion choices and other ways to modify application code for robustness \cite{DBLP:journals/is/AlomariFR14}. Further work in this line showed a sufficient condition to prove robustness for Read Committed Isolation \cite{DBLP:conf/aiccsa/AlomariF15}. 
%
%
Recent research has introduced the concept of split schedules, establishing necessary and sufficient conditions for ensuring transaction robustness across various isolation levels~\cite{DBLP:journals/tods/KetsmanKNV22,vldbpaper,DBLP:conf/pods/VandevoortK0N22,DBLP:journals/sigmod/KetsmanKNV22,DBLP:conf/pods/VandevoortKN23}.
However, a proof technique utilizing split-schedules does not always lead to an efficient algorithm: under a lock-based semantics robustness testing can become coNP-complete~\cite{DBLP:conf/pods/Ketsman0NV20} or even undecidable~\cite{DBLP:conf/icdt/VandevoortK0N22}.
Other work examines robustness within a framework that declaratively specifies different isolation levels in a uniform manner~\cite{DBLP:conf/concur/Cerone0G15,DBLP:conf/concur/0002G16,Cerone:2018:ASI:3184466.3152396}, relying on the key assumption of atomic visibility, which ensures that either all or none of a transaction's updates are visible to other transactions.

Fekete \cite{DBLP:conf/pods/Fekete05} introduced the allocation question: choosing the concurrency control for each transaction separately from a set of available isolation mechanisms; this paper dealt with choosing either 2-phase locking, or snapshot isolation. Recently, other combinations of choices have been considered, such as {combinations of read committed, snapshot isolation and/or serializable snapshot isolation~\cite{DBLP:conf/pods/VandevoortKN23}}. In \cite{DBLP:journals/pacmmod/VandevoortKN24}, the allocation problem is studied in the context of view- rather than conflict-serializability, and is observed that 
for the isolation levels of PostgreSQL, both problems coincide. 

Beyond handling specific transactions that access explicitly identified items, practical scenarios often require working with application code, where the accessed items may vary at runtime. This variability can arise, for example, from values retrieved in earlier queries or user-provided parameters. 
Various approaches have been explored to represent such applications and reason about the different explicit transactions they may generate during execution. The concept of a transaction program was introduced in \cite{DBLP:journals/tods/FeketeLOOS05}, while the abstraction of a transaction template, which we adopt here, was presented in~\cite{vldbpaper}.
%
\new{Adding constraints such as foreign keys quickly makes robustness undecidable~\cite{DBLP:conf/icdt/VandevoortK0N22}, though restricted classes of constraints still allow decidability.}
Vandevoort et al.~\cite{DBLP:conf/edbt/VandevoortK0N23} explored a more expressive variant of transaction programs, offering a sufficient condition for ensuring robustness under read committed isolation. This approach, which relies on a formalism for transaction programs to define potential workloads, stands in contrast to methods like IsoDiff~\cite{DBLP:journals/pvldb/GanRRB020}, where transactions are derived from concrete execution traces.

\section{Conclusions}
\label{sec:discussion}

We introduced a novel optimization method, 
which can enhance performance without requiring modifications to the database system's internals. An evaluation on SmallBank demonstrates that \nameOfApproach can achieve throughput comparable to the unsafe yet default RC isolation level used by some platforms, while maintaining safety. Additionally, it can double throughput compared to executing all under the serializable isolation level.

Our approach builds on the theoretical framework for solving the mixed allocation problem~\cite{DBLP:conf/pods/VandevoortKN23}, which we extend in this paper from fully specified transactions to transaction templates. However, this extension brings several limitations that are not yet addressed in the current formalization. First, it does not account for data dependencies such as foreign keys which more accurately model transactions that can effectively occur. Second, it lacks support for transaction programs that include control structures (e.g., loops and conditionals) as well as operations like inserts, deletes, and predicate reads—scenarios that can trigger the phantom problem. \new{To address these challenges, we can build on ideas from~\cite{DBLP:conf/icdt/VandevoortK0N22, DBLP:conf/edbt/VandevoortK0N23}, though substantial theoretical work is still needed—making this a key direction for future research.}

\begin{acks}
This work was partly funded by FWO-grant G019921. Stijn Vansummeren was partially supported by the Bijzonder Onderzoeksfonds (BOF)
of Hasselt University (Belgium) under Grant No. BOF20ZAP02. The resources and services used in this work were provided by the VSC (Flemish Supercomputer Center), funded by the Research Foundation - Flanders (FWO) and the Flemish Government.
\end{acks}


\bibliographystyle{ACM-Reference-Format}
\bibliography{references}

\inFullVersion{
\appendix
\onecolumn
\section{Appendix}

\subsection{SmallBank Benchmark SQL Code}
\label{sec:app:smallbank}
{
\footnotesize
\begin{minipage}[t]{\textwidth/2-2ex}
\begin{verbatim}
Balance(N):
    SELECT CustomerId INTO :x
      FROM Account
     WHERE Name=:N;
        
    SELECT Balance INTO :a 
      FROM Savings
     WHERE CustomerId=:x;
     
    SELECT Balance + :a 
      FROM Checking
     WHERE CustomerId=:x;
    COMMIT; 

Amalgamate(N1,N2):
    SELECT CustomerId INTO :x1
      FROM Account
     WHERE Name=:N1;
    
    SELECT CustomerId INTO :x2
      FROM Account
     WHERE Name=:N2;
    
    UPDATE Savings AS new
       SET Balance = 0
      FROM Savings AS old
     WHERE new.CustomerId=:x1
           AND old.CustomerId
           = new.CustomerId
    RETURNING old.Balance INTO :a;
    
    UPDATE Checking AS new
       SET Balance = 0
      FROM Checking AS old
     WHERE new.CustomerId=:x1
           AND old.CustomerId
           = new.CustomerId
    RETURNING old.Balance INTO :b;
    
    UPDATE Checking
       SET Balance = Balance + :a + :b
     WHERE CustomerId=:x2;

DepositChecking(N,V):
    SELECT CustomerId INTO :x
      FROM Account
     WHERE Name=:N;
     
    UPDATE Checking
       SET Balance = Balance + :V 
     WHERE CustomerId=:x;
    COMMIT;
\end{verbatim}
\end{minipage}   
\begin{minipage}[t]{\textwidth/2}
\begin{verbatim}
TransactSavings(N,V):
    SELECT CustomerId INTO :x
      FROM Account
     WHERE Name=:N;
        
    UPDATE Savings
       SET Balance = Balance + :V 
     WHERE CustomerId=:x;
    COMMIT;

WriteCheck(N,V):
    SELECT CustomerId INTO :x
      FROM Account
     WHERE Name=:N;
        
    SELECT Balance INTO :a 
      FROM Savings
     WHERE CustomerId=:x;
     
    SELECT Balance INTO :b 
      FROM Checking
     WHERE CustomerId=:x;
     
    IF (:a + :b) < :V THEN 
        UPDATE Checking
           SET Balance = Balance - (:V+1) 
         WHERE CustomerId=:x;
    ELSE
        UPDATE Checking
           SET Balance = Balance - :V
         WHERE CustomerId=:x;
    END IF;
    COMMIT;

GoPremium(N):
    UPDATE Account
       SET IsPremium = TRUE
     WHERE Name=:N
    RETURNING CustomerId INTO :x;
    
    SELECT InterestRate INTO :a 
      FROM Savings
     WHERE CustomerId=:x;
     
    :rate = computePremiumRate(:x,:a);
     
    UPDATE Savings
       SET InterestRate = :rate
     WHERE CustomerId=:x;
    COMMIT;
\end{verbatim}
\end{minipage}
}

}

\end{document}